\documentclass[10pt,journal,compsoc]{IEEEtran}
\usepackage{graphicx} 
\usepackage{algorithm}
\usepackage{tabularx}
\usepackage{algpseudocode}
\usepackage{amsmath} 
\usepackage{tabularx}
\usepackage[T1]{fontenc}
\usepackage{textcomp}
\usepackage{pdflscape}
\usepackage{listings}
\usepackage{xcolor}
\usepackage{float}
\usepackage{adjustbox}
\usepackage{array}
\usepackage{booktabs}
\usepackage{verbatim}
\usepackage{caption}
\usepackage{url}
\usepackage{svg}
\usepackage{hyperref}\hypersetup{pdfborder={0 0 0}}
\usepackage{paralist}
\usepackage{microtype} 
\usepackage{xspace}
\usepackage{rotating}

\newcommand{\pszcusz}{{\scshape pSZ/cuSZ}}
\newcommand{\zfp}{\textsc{zfp}\xspace}

\usepackage{subfigure}

\begin{document}

\title{Lossy Compression of Scientific Data: Applications Constrains and Requirements\\ 
\Large\textit{NSF FZ Project Sarasota Workshop Report}}
\author{
Franck Cappello$^{1,2}$, Allison Baker$^{3}$, Ebru Bozda\u{g}$^{4}$, Martin Burtscher${^5}$,  Kyle Chard$^{2}$, Sheng Di$^{1,2}$, Paul Christopher O'Grady$^{6}$, Peng Jiang$^{7}$, Shaomeng Li$^{3}$, Erik Lindahl $^{8}$, Peter Lindstrom $^{9}$, Magnus Lundborg$^{10}$, Kai Zhao$^{11}$, Xin Liang$^{12}$, Masaru Nagaso$^{4}$, Kento Sato$^{13}$, Amarjit Singh$^{13}$, Seung Woo Son$^{14}$, Dingwen Tao$^{15}$, Jiannan Tian$^{15}$, Robert Underwood$^{1,2}$, Kazutomo Yoshii$^{1}$, Danylo Lykov$^{1}$, Yuri Alexeev$^{1}$, Kyle Gerard Felker$^{1}$\\
\textit{
$^{1}$ Argonne National Laboratory, Lemont, IL, USA.
$^{2}$ University of Chicago, Chicago, IL, USA.
$^{3}$ NSF National Center for Atmospheric Research, Boulder, CO, USA.
$^{4}$ Colorado School of Mines, Golden, CO, USA.
$^{5}$ Texas State University, San Marcos, TX, USA.
$^{6}$ Stanford University, Stanford, CA, USA.
$^{7}$ University of Iowa, Iowa City, IA, USA.
$^{8}$ Stockholm University, Stockholm, Sweden.
$^{9}$ Lawrence Livermore National Laboratory, Livermore, CA, USA.
$^{10}$ KTH Department of Applied Physics, Solna, Sweden.
$^{11}$ Florida State University, Tallahassee, FL, USA.
$^{12}$ University of Kentucky, Lexington, KY, USA.
$^{13}$ RIKEN R-CCS, Japan.
$^{14}$ UMass Lowell, Lowell, MA, USA.
$^{15}$ Indiana University Bloomington, Bloomington, IN, USA.}
}

\date{March 21 2025}

\maketitle



\setcounter{tocdepth}{4}
\tableofcontents{}
\pagebreak

\section{Introduction}
Scientific simulations, experiments, and observations are producing increasing volumes of data due to the change of supercomputer generation (from petascale to exascale) and the update of large scientific instruments (accelerators, light sources, telescopes). In many situations, the produced data is too large to be communicated on a network, stored in storage systems, and analyzed with user tools. The scientific community's response to this challenge is scientific data reduction. Reduction can take many forms, such as triggering, sampling, filtering, quantization, and dimensionality reduction. This report focuses on a specific technique: lossy compression. Compared with other scientific data reduction techniques, lossy compression keeps all data points. It leverages the correlations between data points and the reduction of data point accuracy to reduce the scientific data. To preserve the same opportunities for scientific discoveries from lossy compressed data as from noncompressed data, compression techniques need to respect user quality constraints that generally concern the preservation of quantities of interest (QoIs) to a certain accuracy. In addition, in order to be useful, a lossy compression technique needs to satisfy user requirements in terms of compression ratio (by what factor the data has been reduced compared with the original version) and compression speed (how fast and at what throughput the scientific data can be compressed). 

While many papers have been published on lossy compression techniques and reference datasets are shared by the community \cite{sdrbench}, there is a lack of detailed specifications of application needs that can guide the lossy compression researchers and developers. This report fills this gap by reporting on the requirements and constraints of nine scientific applications covering a large spectrum of domains (climate, combustion, cosmology, fusion, light sources, molecular dynamics, quantum circuit simulation, seismology, system logs). For every application, the report details the motivations for compression, the current uses of compression, the compression requirements, the analysis and quality requirements, the performance requirements, the constraints on sustainability, integration and installation, the special needs, and the expected changes. Table \ref{tab:appfeatures} summarize these needs for the nine applications.

To complement the specifications of application needs, the report details the main lossy compression technologies at the time of the writing: SZ, ZFP, MGARD, LC, SPERR, DCTZ, TEZip, and LibPressio. The report presents the history, principles,  method for error control, hardware support, unique features, and impact of every compression technology. Presenting the application needs and the existing compression technologies in one report allows readers to understand how the current compression technologies respond to the application's needs. The report discusses the existing gaps between the application needs and the lossy compression technology capabilities in the application sections. The co-authors of this report hope that this report will inspire new research to fill existing gaps.

\subsection*{Data collection method} 
The data for this report was collected over three days during the NSF FZ project meeting at Sarasota in February 2025. Experts from the nine application domains presented their applications and the constraints and requirements regarding lossy compression. The lossy compression experts presented their technologies in detail. The general presentation of the applications and compression technologies was followed by a one-to-one meeting between the applications experts and lossy compression experts to refine the specification of the requirements and constraints and to identify needs that were not expressed during the application presentation. The current report was written from these interactions and in collaboration with the application and lossy compression technology experts. This report reflects the state-of-the-art applications in March 2024 and the lossy compression technologies in January 2025. The co-authors of this report will update it as needed to reflect changes.

\subsection*{Report organization}
The report is organized into four main sections. The first section is this introduction. Section 2 details the nine applications and their lossy compression needs. Section 3 presents the lossy compression technologies. Section 4 summarizes the gap analysis, and Section 5 briefly summarizes the conclusions.

\section{Applications} 




\subsection{Climate}
Understanding the Earth's climate system has long been of interest, particularly as a requirement for better predicting future climate states.  Climate simulation models have become increasingly complex over the decades as computing resources have grown in power and sophistication  (\cite{washington2005introduction, gettlemanbook}). 
Modern Earth system models (ESMs) are widely used to study past, present, and future climate states; and better understanding our changing climate has become an urgent priority for society.  
ESM simulations are well known for producing enormous amounts of output data, as increases in computational power have enabled finer spatial and temporal resolutions, longer simulations, and larger ensembles.  
And while these advances are desirable for more accurate and realistic simulations, the associated data storage requirements are often prohibitively large, since supercomputing storage capacities have not increased as rapidly as computational power and financial constraints limit the storage capacity available at many institutions \cite{kunkel2014}.

\subsubsection{Motivations for Compression}
The increasing data generation trend in model-based climate research is unsustainable. 
The Coupled Model Intercomparison Projects (CMIPs), which facilitate international ESM comparisons via specific data and experiment specifications, have grown substantially over the years in terms of data volume requirements. The most recent CMIP6 effort resulted in roughly 28 PB from 45 modeling institutes, hosted by the Earth System Grid Federation  \cite{ESGF}, while CMIP5 generated 2 PB of data and CMIP3 generated only 40 TB \cite{cmip6}.  
The contribution of the National Center for Atmospheric Research (NCAR)  to CMIP6  with the Community Earth System Model
(CESM) \cite{cesm2013}, for example, was 2 PB of of postprocessed time series data; and, notably, the vast majority of CMIP runs are not even considered high resolution.
The recent push toward kilometer-scale models (i.e., ``ultra-high" resolution), exemplified by the DYAMOND (DYnamics of the
Atmospheric general circulation Modeled On Non-hydrostatic
Domains) project \cite{Dyamond}, is severely impacted by storage constraints.
In fact, computational and storage costs were so high for the initial DYAMOND atmosphere-only model experiments 
that simulations were limited to 40 days,
and 3D variable output was scant---in some cases outputted 12$\times$ less often compared with 2D data.  The DYAMOND contribution from  SCREAM (Simple Cloud-Resolving E3SM Model \cite{e3sm-scream2023}), run at 3.25 km resolution, was nearly 4.5 TB of data per simulated day \cite{caldwell2021}.
The reality is that climate scientists are often unable to store all of the simulation output that they would like, and this limitation directly impacts climate science research investigations.

\subsubsection{Current Uses of Compression}

Currently the climate community widely utilizes ``standard'' lossless compressors that are widely available, such as Gzip \cite{Deutsch_1996} and, increasingly, Zstd \cite{Facebook_Inc_2019}. Some researchers in this community use simple lossy compression schemes such as bitgrooming \cite{Zender_2016} and digit rounding \cite{Delaunay_Courtois_Gouillon_2018}, which have native support in key I/O libraries such as netCDF and to a lesser extent HDF5; but this usage is not widespread in the community or by major facilities.  Few researchers currently use advanced lossy compressors such as SZ or ZFP \cite{Underwood_Bessac_Di_Cappello_2022,Klower_Razinger_Dominguez_Duben_Palmer_2021}.

\subsubsection{Compression Requirements}

Requirements for lossy data compression have been proposed by many over the years as a means of mitigating the big data storage problem in climate (e.g., see \cite{woodring2011, hubbe2013, bicer13, baker2014, kuhn2016, baker2016, Underwood_Bessac_Di_Cappello_2022}.
While lossy compression is attractive, a number of challenges associated with this effort in practice have hindered the widespread acceptance and use of lossy compression in the Earth system modeling community.

\paragraph{Analysis Metrics and Quality} For ESMs such as the popular CESM, climate scientists are reluctant to lose any information and have concerns about the effects of data compression-induced artifacts given the societal  scrutiny over future climate and possible implications of model predictions.
Additionally, because of the computational power and storage required by most climate simulation models, large datasets are typically generated by research institutions such as NCAR and then made publicly available for the wider climate research community (e.g., \cite{kay2015, lens2, chang2020}. 
This workflow style means that those generating (and possibly compressing) the output data often do not know how the data will be analyzed. Therefore, identifying particular data characteristics to preserve (e.g., extreme values, subtle shifts in seasonal cycles, changes in gradients) so as not to affect scientific analysis is nontrivial, encouraging a cautious approach.  

Packages such as LDCPY \cite{Pinard_Hammerling_Baker_2020} 
provide a step in this direction.
LDCPY offers various metrics, including dSSIM, east-west covariance, standard deviation, probability positive, mean absolute error, deseasonalized lag 1 correlation, annual harmonic mean, z-score for a zero mean, Pearson correlation coefficient, and the p-value for the Kolmogorov--Smirnov test.\footnote{The p-value of the Kolmogorov--Smirnov test may be both oversensitive and undersensitive for this use case \cite{Underwood_Bessac_Di_Cappello_2022}.}  Of these, dSSIM is emerging as a key metric of interest and has received additional study. 

dSSIM is a variant of the classic Structural Similarity Index Metric (SSIM) extensively studied for use with climate data. dSSIM has a few changes relative to SSIM: (1) values are normalized between 0 and 1 by using a linear scaling and then quantized to an integer in the range of [0,255], which  corrects for large value ranges and represents the conversion to a colormap in 8-bit color space; (2) the standard values of $k_1$ and $k_2$ are replaced with $1\times10^{-8}$ to better reflect its use for data and not iamges; and (3) it uses a Gaussian convolution kernel that preserves NaN values and uses \texttt{fill} boundary semantics that better handles NaN's used to represent missing values and values near the edges of an image.
This produces a value between 0 (poor) and 1 (perfect), which is compared with a desired similarity threshold of 0.99919 or 0.995 for more ``aggressive data compression.''

\paragraph{Performance Requirements} The time and ease of reading data for analysis are extremely important to climate scientists. 
ESMs such as CESM need fast and parallel I/O for netCDF files, as well as fast-enough support for a wide variety of tools used by the climate community to analyze data. These requirements can inhibit the use of some lossy compressor technologies, particularly if used to compress when the data is initially output (as opposed to in a postprocessing step).  
If possible, climate researchers hope to obtain a 2--3$\times$ improvement over lossless compression ratios while preserving the scientific integrity of their data with comparable bandwidth to using lossless compression on CPUs.
Preliminary results from \cite{Underwood_Bessac_Di_Cappello_2022, Klower_Razinger_Dominguez_Duben_Palmer_2021} suggest that this is possible at least for some fields in some datasets but needs a more comprehensive validation with many fields and datasets which in turn requires more scalable validation methods.

\paragraph{Sustainability} The required lifetime of the data is  long (possibly ``indefinite''), and the number of users accessing the data is very large.  This requirement arises from the need to compare climate predictions over time.  In terms of suitable lossy compressors, this aspect means that the climate modeling community is hesitant to use compressor technologies that may not be available/supported for the long haul (rendering old data unreadable) or that may require any additional burden on the users in terms of reading the compressed data.

\paragraph{Integrations and Installations}
The climate community has a wide range of tools used by different segments of the community, including  PnetCDF, HDF5, Python, Julia, and the NCAR command language.
These tools gain access to the software through a variety of methods.
Many users rely on sitewide deployments in software modules at NCAR, but increasingly package managers are used by users to provide their own libraries, such as Anaconda and pip for Python. In addition,  more general HPC-focused package managers such as Spack are used.
An approach that enables compatibility and feature parity across all  these methods of accessing data is critical to adoption by facilities.

\paragraph{Special Needs: Automated Configuration and Uncorrelated Dimensions} Evaluating the information loss for climate application given the vast diversity of climate data fields is nontrivial, requiring automated approaches to scale analysis and determine configurations.  ESMs typically output hundreds (or even thousands) of variables with very different characteristics, as well as spatial and temporal dependencies that require the individual treatment
of variables and not a ``one-size-fits-all” approach \cite{baker2014, baker2016, poppick20}.   Indeed, typical compression metrics (e.g., RMSE, PSNR) are simply not able to capture all the kinds of compression artifacts that might be important in a wide
variety of analysis settings (e.g., see \cite{baker2014, baker2016, baker2017, Hammerling2019, poppick20, baker2019}).

Additionally, some climate datasets contain unique features that make compression of specifically uncorrelated dimensions more challenging.
Normally, compressors assume that all dimensions (e.g., latitude, longitude, height) of data are correlated; but in some climate datasets, the data may be correlated in some dimensions
 (e.g., latitude or longitude) but not all (e.g., height) because of limits in resolution or natural phenomena.
This situation  affects compression because the uncorrelated dimensions can cause mispredictions in prediction-based compressors, resulting in lower compression ratios.
Compressors such as CLISz \cite{Jian_Di_Liu_Zhao_Liang_Xu_Underwood_Wu_Huang_Chen_et_al._2024} provide a step in addressing this issue by automatically detecting uncorrelated dimensions and not predicting across them.

\paragraph{Expected Changes}
In the future,  more and more climate codes are expected to  migrate from CPU to GPU; and, with that migration, the need to compress on the GPU will increase in order to avoid unnecessary costs of copying data between CPU and GPU.
Additionally, currently most climate data takes the form of structured grids. But more and more climate codes  are moving from structured  to unstructured grid formats such as MPAS-A \cite{Climate_Ocean_and_Sea_Ice_Modeling_COSIM_Team_at_Los_Alamos_National_Laboratory_The_National_Center_for_Atmospheric_Research_2015}.  This trend is not expected to decrease the need for storage and in turn data compression, but it will require updates to analysis codes and corresponding compression pipelines to support unstructured grids, which are currently supported only  by some of the most recently developed lossy compressors \cite{Ainsworth_Tugluk_Whitney_Klasky_2020,Di_Liu_Zhao_Liang_Underwood_Zhang_Shah_Huang_Huang_Yu_et_al._2024,Ren_Liang_Guo_2024,Wang_Pulido_Grosset_Jin_Tian_Zhao_Ahrens_Tao_2024}.

\subsection{Combustion}

While turbulent fluid motion is a common thread through computational fluid dynamics (CFD) applications, the multiphysics coupled with fluid motion spans many different subdisciplines, including chemistry in the gas phase and at surface interfaces, plasma physics critical to  energy-efficient chemical manufacturing and fusion energy, aerosol growth and coagulation, and spray atomization and evaporation. CFD at the exascale on DOE leadership-class supercomputers runs on thousands of computational nodes powered by GPUs and generates massive volumes of primary data, requiring storage and analysis of quantities of interest (QoIs). It is infeasible to store data at sufficient frequencies to capture highly intermittent phenomena that occur in these transient simulations.

\subsubsection{Motivations for Compression}

The datasets produced by exascale direct numerical simulation (DNS) codes are typically 2--3 terabytes per checkpoint file, with approximately 500 checkpoint files saved to storage to track the temporal evolution. Each checkpoint contains between 25 and 150 dependent variables per grid point, comprising density, three components of momentum, total energy, and species mass fractions. The computational mesh contains nominally upwards of 10--20 billion grid points, and the DNSs were performed  on 2,000 nodes on Frontier at the Oak Ridge Leadership Computing Facility (1/4 of the machine). The runs spanned approximately 100,000 timesteps and required tens of millions of GPU node-hours to complete. Owing to their enormous cost, trustworthy machine-learning-based data reduction is essential to ensure the data's downstream utility within the CFD community~\cite{cai2022physics, brunton2020machine, Duraisamy2019, Beck2021}. Researchers from Sandia National Laboratories and collaborators have created BlastNet \cite{chung2022aecs}, a public Web-based repository for 3D compressible turbulent flow DNS datasets adhering to FAIR principles\cite{WilkinsonFAIR}. This wide range of applications and datasets provides the breadth of requirements for various reduction models.

\subsubsection{Current Uses of Compression}

Scientists have been using both lossless and lossy compression techniques to manage the vast volumes of simulation data produced by CFD simulations.
For example, lossless compression is often favored when high accuracy is paramount, since it guarantees that no information is lost.
As noted in several studies, however, the application of lossy compression techniques 
or these datasets remains limited because of the stringent accuracy requirements for both raw data and QoIs in downstream analyses.
For instance, the SZ2 compressor has been employed to compress BlastNet data, an approach that is part of an effort to reduce storage requirements while preserving essential features of the data for further analysis.
SZ has
demonstrated potential in compressing large-scale DNS simulation data without sacrificing critical statistical properties, but it remains largely underexplored in broader contexts, especially given the need to preserve the integrity of topological features and high-dimensional data across various scales.

\subsubsection{Compression Requirements}
\paragraph{Analysis Metrics and Quality}
%
We identified two combustion applications that have diverse compression needs: S3D simulations and BlastNet machine learning models.  The simulation needs are driven mainly by checkpointing and in situ analysis, while BlastNet focuses primarily on super-resolution. For S3D simulations, the primary need for compression is to reduce the storage size while saving events of interests that happen during the simulation.  This requires high-throughput GPU compressors because (1) S3D's simulation outputs are immediately available on GPUs, (2) S3D requires fast checkpoint/restart, and (3) snapshots for backward analysis must be taken on the fly, requiring high-throughput compression.  

Table~\ref{tab:appfeatures} summarizes the features of the primary CFD data ranging from topological feature descriptors to statistics. 
For example, merge trees characterize topological changes in turbulence structures as runs are going~\cite{chong1990}.  Merge trees are also used to steer analysis and checkpoint, based on persistence-based simplification of merge trees to filter out unimportant branches.  Morse---Smale complexes are another descriptor that contain critical points (minima/maxima/saddles) and their relationships; they lead to topological segmentations that help characdterize important regions such as burning cells. Discontinuotiies (e.g., shocks) need to be preserved in compression because the shock structures have critical points around them in a (wide) stencil; downstream analyses are applied to regions with discontinuities. Another example is the \emph{level-set restricted Voronoi decomposition}~\cite{neuroth2022level} of the domain; this descriptor captures voxel-level information of each Voronoi cell and performs a high-level analysis of the cell representation.  The Voronoi decomposition is also important for understanding how the distribution/aggregation of different variables correlates with the physics.  

The region of interest (RoI) relevant to reacting CFD for combustion needs to capture the strong coupling between the flame wrinkling and the underlying turbulent strain, where the flame response is characterized in composition phase space (comprising species concentrations and temperature, which control the burning rate and are modulated by turbulent strain), while turbulent coherent motions occur in physical space. Therefore, the ability to go back and forth between physical and composition space is needed to understand causality.  
For composition space, the distance function (comprising isosurfaces removed from the flame location) can be used to identify regions upstream of the flame sheet toward the reactants' preheat zone and regions downstream toward the burnt products zone.  The effect of turbulent strain on the flame structure is analyzed by obtaining statistical means of scalar quantities (e.g., temperature or species concentrations) conditioned on the distance from the flame surface. The level sets and distance function within the flame brush should be faithfully captured in any lossy reduction scheme with stringent error bounds of $\mathcal{O}(10^{-3})$ to $\mathcal{O}(10^{-4})$ and with more relaxed error bounds outside of the flame brush.

\begin{table*}[hbtp]
\centering
{\centering\footnotesize
\begin{tabular}{|p{1.9cm}|p{4.25cm}|p{4.25cm}|p{4.25cm}  |}
\hline
 Attribute & Combustion & Aerosols in Climate Models & Hypersonics\\
\hline
  \hline
 Tensor Correlations & gas phase species reaction rates & black carbon reaction rates & dissociative air and ablative material reaction rates \\
 \hline
 Scalar QoI & concentrations, temperature, progress variable, scalar dissipation rates, mixture fraction   &  concentrations, temperature, aerosol population balance, scalar dissipation rates   & turbulent Mach number, multi-temperature, mass fractions, dilatation, detonation speed\\
 \hline
 Vector QoI & velocity, vorticity, rate of deformation tensor, scalar gradients & velocity, vorticity, rate of deformation tensor, scalar gradients&  velocity, vorticity, rate of deformation tensor, scalar gradients\\
 \hline
 Nonlinear QoI    &gas-phase thermal reaction rates for ammonia-hydrogen and sustainable aviation fuels & climate reaction rates, aerosol coagulation, agglomeration and oxidation rates & compressible turbulent kinetic energy budget, thermal and nonthermal reactions \\
 \hline
 Error Bounds & \multicolumn{3}{|c|}{Normalized Root Mean Squared Error (NRMSE): Stringent $\mathcal{O}(10^{-4})$, Relaxed $\mathcal{O}(10^{-3})$}\\
 \hline
Level Sets and ROI Detection & flow topology, level sets conditional on reaction progress and mixture fraction & level sets conditional on aerosol surface reactions  and  morphology
 & level set conditional on reaction progress and normalized speed of detonation wave\\
 \hline
\end{tabular}
}
\vspace*{0.5cm}
 \caption{Examples of Application Reduction-Related Features of Three Diverse Classes of CFD Applications} 
 \label{tab:appfeatures}
\end{table*}

\paragraph{Performance Requirements}

This application does not have a strict requirement for compression ratio but rather cares about accuracy in terms of both raw data values and QoIs.  Driven by the needs of very high data accuracy, a compression ratio with 5--10$\times$ (or even 2--3$\times$) is already useful for the application.  For S3D simulations high-throughput compression on GPUs is required because S3D data is already on GPUs during the simulation; on the contrary, BlastNet could afford longer compression time, yielding a higher compression ratio.

\paragraph{Sustainability}

Compressed datasets need to be retained for at least 5--10 years to support long-term scientific analysis and validation. Given the large-scale nature of combustion simulations, which generate multiterabyte snapshots, maintaining accessibility to stored data is crucial for future studies, comparative analyses, and potential reprocessing with improved methodologies. Storage constraints drive the need for efficient compression strategies that balance high compression ratios with the preservation of key statistical and topological features. Additionally, ensuring compatibility with evolving software frameworks and maintaining support for partial decompression and random access further contribute to the long-term sustainability of these datasets.

\paragraph{Integrations and Installations}

Compression tools are integrated into workflows primarily through C++ and Python, with a strong preference for HDF5 and ongoing collaborations involving ADIOS2. In combustion simulations, compressors are typically invoked as part of custom-built simulation frameworks, whereas analysis tools may leverage package managers such as pip or conda for easier installation. Currently, SZ2 is the primary compression framework in use, although alternative approaches are being explored based on accuracy and performance requirements. While speed is not a primary concern for offline compression, maintaining data fidelity is critical, particularly for preserving statistical properties and topological features.

\paragraph{Special Needs: Toplogy Preservation}

A key requirement for analysis is the ability to preserve topological and structural features critical to understanding turbulence and combustion dynamics. In particular, Morse--Smale complexes are essential for capturing critical points and their hierarchical relationships, which help characterize complex flow structures. The ability to simplify merge trees through persistence filtering is also necessary to remove insignificant branches and focus on meaningful topological changes during simulations. Additionally, maintaining spatial discontinuities, such as shocks, is crucial, since these regions contain critical points that influence combustion behavior. Compression strategies must ensure that statistical properties, including point statistics, joint probabilty density functions, and gradient-based metrics, remain accurate. Moreover, partial decompression and random access capabilities are needed to facilitate efficient analysis of large datasets without requiring full decompression.

\paragraph{Expected Changes}

The core requirements for accuracy-driven compression in combustion simulations and machine learning applications are expected to remain stable, but several evolving factors may influence future needs. As simulations scale up, with terabytes per snapshot anticipated on systems such as Frontier, efficient compression strategies will become even more critical for managing storage and I/O constraints. Additionally, there is ongoing exploration of alternative compression frameworks beyond SZ2 to improve accuracy and efficiency. The integration of in situ analysis and real-time event detection may introduce new demands for high-throughput compression on GPUs to support on-the-fly processing. Advances in Morse--Smale complex analysis and topological tracking across ranks could also shape future requirements for preserving hierarchical structures and turbulence features. While current interfaces, such as C++/Python with HDF5 and ADIOS2, are well established, future optimizations may focus on enhancing random access and partial decompression to improve usability for large-scale scientific analysis.

\subsection{Cosmology}
Cosmology, the study of the Universe on its largest scales and across its entire history, explores some of the most exciting questions in fundamental physics: the nature of dark energy and dark matter, the origin of primordial fluctuations, the origin and evolution of galaxies, and the intergalactic medium. Interpreting the ongoing and future sky surveys involves solving an inverse problem: deducing underlying physics from observational data.  Here, the numerical simulations play an essential role as a forward model, since they are the only accurate way to model the nonlinear evolution of the Universe.

\subsubsection{Motivations for Compression}

Cosmological simulations impose significant challenges in terms of data management due to the sheer volume being generated and processed. These simulations typically involve modeling the evolution of hundreds of billions, even trillions, of particles or cells.  The output datasets therefore reach sizes ranging into petabytes, necessitating efficient compression techniques to reduce storage and I/O bandwidth.  Effective compression methods not only minimize storage requirements but also facilitate quicker data access and analysis.  However, achieving good compression ratios while preserving scientific fidelity poses a delicate balance, requiring assurance that the compressed dataset's crucial details of cosmic structure and dynamics are accurately preserved.

To give a real-life example, in a current INCITE project a hydrodynamics simulation mixed with a 12,288$^3$-element N-body array and hydrodynamics was run on NERSC's Cori, with storage requirements holding back the Nyx code from facilitating an 16,384$^3$-element simulation. Even with a double-precision array of 12,288$^3$ elements, 13.5 TB of storage is allocated, with 14 similar arrays used for a checkpoint.

\subsubsection{Current Uses of Compression}

Lossless compression methods have been deterministically tested for the Nyx simulation, achieving a compression ratio of 2 to 3$\times$~\cite{use-case-Franck}, far below the 10$\times$ or higher needed to address the storage challenges while maintaining data quality for post hoc analysis. In recent years, early explorations of integrating lossy compression into the simulation workflow have been actively pursued by leveraging CPU versions of SZ2 and SZ3 for AMReX, the basic data structure used for the data interpretation during simulation. .

Recent studies in error-bounded lossy compression, particularly SZ-based methods such as TAC+, optimize data reduction for adaptive mesh refinement (AMR) and cosmology simulations by leveraging the nature of AMR data: multiresolution, adaptive partitioning, and therefore shared encoding can be used to improve efficiency while preserving accuracy \cite{tac+}. Integrating lossy compression with visualization workflows further enhances storage efficiency and minimizes artifacts \cite{wang2024multires}, while in situ solutions address real-time data reduction challenges \cite{wang2023analyzingimpactdatareduction}. These innovations significantly reduce I/O overhead and improve postanalysis for large-scale simulations.

\subsubsection{Compression Requirements}

we list below the requirements 
for integrating the compression pipeline into the cosmology simulation in order to address the storage challenge.
Included are
the case-by-case needs for post hoc analyses in terms of metrics and quality and the basic throughput requirements. In addition, since the internal data is based on AMR, the compression, while taking advantage of it to adjust error tolerance, needs to be suitable for this data format.

\paragraph{Analysis Metrics and Quality}
Currently, the following aspects of the simulation are considered: halos (halo-finding is one of the QoIs), power spectrum, PDF, Gimlet, and $k$-point function. These analysis metrics can be consistently the same in the following five years, and the simulations generate deterministic results to analyze. The specific compression quality metric is case by case. For the power spectrum, introduced errors must be bias-free, indicating the desire for uniformly random error. SZ may incur patterned errors, however, invalidating the randomness, which could require special treatment for the cosmology application. On the other hand, because the simulation will be conducted using more levels of refinement, it is worth studying how different refinement levels can tolerate the introduced compression error.
In addition, data integrity is preferred when considering the blockwise treatment during compression for data-parallel applications; a series of studies on data granularity have been done recently \cite{wang2023americ,tac+,wang2024multires}.

\paragraph{Performance Requirements}  Compression ratio is the most significant factor influencing the choice of compression solution. Specifically, ratios exceeding 10$\times$ need to be guaranteed, and approaching 1--2 orders of magnitude of the data reduction rate is desired. 
Currently, SZ can deliver the desired compression ratio even with multimodal data formats: Regular-grids can be handled by generic SZ3, and the runtime AMR data can be handled by specialized compression techniques such as AMRIC \cite{wang2023americ} and MultiReZ \cite{wang2024multires}.

Speed considerations are important and are twofold: (1) compression must not cause significant delays in I/O operations, and (2) decompression performance should not hinder post hoc analysis workflows. While higher computational costs are still affordable for decompression, minimizing the significant bandwidth constraints swiftly is prioritized.
Currently, along with the parallel I/O (e.g., studied in \cite{wang2023americ}), CPU-based compression is still beneficial, although faster compression as data processing always can be desired. The long-term concerns involve the shift to more heterogeneous compute paradigms in ExaSky cosmology applications: the Hardware/Hybrid Accelerated Cosmology Code (HACC) and Nyx,a cosmological simulation code, are readying the GPU contribution to the compute capabilities. Fortunately, the compression research with heterogeneous computing preparation has been ongoing for years; the resident data on GPU memory can be processed in situ utilizing the GPU-based compressors (e.g., \textsc{pSZ/cuSZ}, cuSZp, FZ-GPU) to avoid extra memory copy.

\paragraph{Sustainability}
Data needs to be preserved for post hoc analysis. The raw data is still sampled and partially open-accessible. As the simulation scales up, however, extensive storage of full-scale checkpoints is unsustainable. For example, the code base supports Nyx running at a data scale of a double-precision $16,384^3$ for N-body hydrodynamics simulation. However, larger runs have instead used $12,288^3$ because of the limited storage capacity for checkpointing. Additionally, the longevity of data preservation is yet to be determined based on the specific exploration type from the simulation.

\paragraph{Installations and Integrations} 
C/C++ and HDF5 are used primarily for invoking compressors because of their performance and compatibility with the workflows. For post hoc analysis, Python may also be utilized and is particularly beneficial for the flexible capabilities of partial decompression.

On-the-fly GPU compression with HDF5 filter support is preferred. This is particularly important when designing the compression pipeline, 
because the ability to perform partial decompression would be highly beneficial and can significantly enhance productivity for domain scientists. 
However, HDF5 H5Z filters have limited interoperability and restricted access to the internal workings of each compressor. 
Partial decompression still needs to be addressed at the compressor level, necessitating the continued development of proper buffering mechanisms in response to this requirement.

\paragraph{Special Needs: Portable Compression/Decompression} In addition, support for multiple GPU backends is required, necessitating portability solutions.
For example, compression and decompression on different machines and hardware must be consistently reproducible. For instance, for {\pszcusz} GPU compression on CUDA GPU systems, it must be compatible with decompression on CPU or decompression on AMD GPU systems.

\paragraph{Expected Needs}
With the volume of data from simulation, random-access featured exploration is expected to become increasingly important and challenging. Domain scientists want swift data access while conducting postanalysis. One representative operation over the data is slicing, which is far more complicated than it appears considering the multidimensional nature of data.  HDF5 data format solves this with chunking; but if chunks are not well aligned to post hoc analysis access patterns, performance will suffer.

\subsection{Magnetic Confinement Fusion }
Fusion energy holds the promise of a clean, baseload generation source of electricity in a decarbonized future. Magnetic confinement is one approach for achieving viable fusion energy, and tokamaks are the predominant experimental direction for magnetic confinement fusion today. The ITER tokamak, currently under construction in France, aims to demonstrate technical feasibility of a burning plasma with a tenfold ($Q\geq10$) power gain.

\subsubsection{Motivations for Compression}
Experimental tokamaks, although smaller than ITER, generate vast quantities of data through their extensive instrumentation. These datasets are multimodal and can accumulate over years of research campaigns, making data management a significant challenge. One major data source is electron cyclotron emission imaging (ECEi), used in tokamaks such as DIII-D in San Diego, CA. ECEi captures snapshots of electron temperature fluctuations at a high temporal frequency of 1 MHz and a spatial resolution of $20\times8$ grid points. The sheer volume of data generated by ECEi, especially considering that measurements are continuous and span extended periods, results in large datasets that can be difficult to store and manage without compression.  In addition to diagnostic data, first-principles simulations of plasma physics, such as high-resolution gyrokinetic models, produce large output files that are critical for understanding and predicting fusion device performance. These simulations, run on leadership-class high-performance computing resources, also generate significant disk storage requirements, often reaching terabytes per simulation. The combination of high-frequency diagnostic data and high-resolution simulation outputs creates immense data volumes that complicate scientific analysis.

To facilitate the use of fusion data in machine learning models for predicting fusion disruptions, scientists downsample the high-frequency ECEi data before transferring it from cold storage for model training. Compression is essential to reduce the storage footprint and ease data transfer bottlenecks, while maintaining key features necessary for training the machine learning models. However, ensuring that compression does not degrade important temporal or spatial information remains a critical challenge in this context.

\subsubsection{Current Uses of Compression}

With the unique characteristic of ECEi data usage, that is, retrieving data from cold storage for model training, the primary uses of compression are twofold: (1) transferring data from cold storage to supercomputer centers and (2) using compressed data for training machine learning models. To this end, scientists are currently using SZ compressors via HDF5 filters.   These compressors provide high compression ratios while preserving sufficient data quality for machine learning applications, allowing for more efficient storage, transfer, and processing of large datasets.


\subsubsection{Compression Requirements}

Researchers have been investigating the data management needs and compression requirements in fusion research, with a focus on high-temporal-resolution diagnostic data, such as ECEi. For example, Churchill et al.~\cite{churchill2019deep} explore the use of deep convolutional neural networks to analyze high-temporal-resolution ECEi data from the DIII-D tokamak, emphasizing the challenges of managing large datasets for machine learning applications.

While the primary quality metric for ECEi is absolute error, scientists care about the machine learning (ML) model accuracy and preservation of spikes/peaks in 1D profiles.  For example, with the current practice of temporal downsampling, interpolating temporal dimensions by 4$\times$ could cause trouble for training; instead, scientists are careful choosing the downsampling ratio with 1 kHz frequency.  Regarding features of interest, unlike fusion simulation data, the low spatial resolution ($20 \times 8$) will not resolve detailed features such as blobs.  However, scientists care about spikes and peaks in the 1D profile.
With the introduction of lossy compression, scientists would desire a compressed reprsentation over the temporal dimension in a transparent manner.  Scinetists would also favor a physics-informed approach for compressing the ECEi data for higher authenticity of ML models.

\paragraph{Performance Requirements}
Scientists anticipate at least 5$\times$ the minimum compression ratio for ECEi datasets.  
For disruption prediction models, it may also be possible to use reduced precision (e.g., single precision instead of double precision) of ECEi time series data for training ML models; but scientists will need to investigate further whether the reduced precision will downgrade model prediction capabilities.

Current compressors can achieve the performance requirement without considering QoI (peaks/spikes) preservation.  Since the original high-temporal-resolution, uncompressed data is permanently archived in cold storage, the speed requirements for this application are moderate, meaning that real-time or near-real-time compression is not currently necessary.  
That said, further preserving peaks/spikes in compression may require GPUs, as demonstrated in recent literature in topology preservation \cite{msz}.
Moreover, as scientists anticipate an automatic and coupled workflow in the future, it would require higher (de)compression performance.  

\paragraph{Sustainability}  

Because all original, non-compressed data is securely stored in cold storage, there is no immediate need to retain compressed versions for the long term. The cold storage ensures that the raw data can always be accessed for future use or reanalysis, making the compressed data more of a temporary, efficient format for model training rather than a permanent archival solution.

\paragraph{Integrations and Installations}
Scientists would expect an easy-to-use and transparent approach to use compression; for example, one would expect an easy installation of compressors using pip installation. 
In this context, transparency refers to the ease with which scientists can use compression tools without needing to understand the underlying algorithms or technical details. The process should be seamless, allowing users to focus on their analysis while the compression tool handles data reduction efficiently in the background.

\paragraph{Special Needs}
Unlike traditional simulations or datasets where QoIs such as error margins or physical properties are well defined, ML applications in fusion rely on more abstract and nuanced QoIs that are tied to the performance of predictive models, such as disruption prediction in fusion reactors. These models require the preservation of implic features, which remains an open challenge and may not be easily measured in traditional terms.

\paragraph{Exepected Changes}

We expect to see increased automation in ML workflows, driving demand for faster and more efficient compression techniques. As data volumes grow with advanced tokamaks and simulations, compression methods will need to handle high-temporal-frequency imaging data while preserving key features such as spikes in 1D profiles. Real-time compression and decompression may become more common, especially for streaming data to supercomputing centers. Additionally, the development of physics-informed compression methods will become crucial to maintain data authenticity for ML model training, with specialized techniques tailored to fusion research.


\subsection{Light Sources}
Light sources such as the Linac Coherent Light Source (LCLS) operated at SLAC National Accelerator Laboratory (SLAC) and the Advanced Photon Source (APS) operated at Argonne National Laboratory  allow scientists to improve their understanding of the materials that make up our Universe, as well as improving our understanding and ability to construct advanced electronics, pharmaceuticals, and nanoscale technologies and to study the makeup of living things.  
While the two facilities have significant differences, they both produce enormous volumes of data and are expected to produce more as updates are completed that will result in dramatically increased X-ray pulse rates.  For LCLS-II the repetition rate will increase to 1 MHz compared with 120 Hz for LCLS-I.

\subsubsection{Motivations for Compression}
Over the next few years the LCLS-II project at SLAC and APS-U at Argonne will deploy new area detectors for these high-rate ultrafast X-ray shots.  At SLAC, the devices with the highest data volume are 16 megapixel area detectors running at 35 kHz, producing $\sim$1 TB/s in just one experimental hutch. Three other hutches will have 4-megapixel detectors running at similar rates.  Argonne is installing devices capable of producing similar data volumes. The storage and bandwidth costs for this data volume are prohibitive, so real-time data reduction is needed to economically store the data produced by this next generation of detectors.

\subsubsection{Current Uses of Compression}
The adoption of compression varies significantly by site.
With existing low-rate (120 Hz) LCLS-I datasets, LCLS has studied the effect of lossy compression in crystallography experiments using a custom RoibinSZ algorithm  \cite{underwoodROIBINSZFastSciencePreserving2023} as well as low-intensity and high-intensity SAXS/WAXS experiments using the SZ3 lossy compression algorithm \cite{liangSZ3ModularFramework2023}.  For the former LCLS has achieved a factor of 90 reduction, while for the latter LCLS has achieved a factor of 9 reduction without compromising the physics results.  LCLS has also begun investigating the LC compression package \cite{burtscherBurtscherLCframework2024}.
At the APS, datasets are compressed with lossless compressors such as ZStandard \cite{zstd} as integrated via HDF5.
Additionally, researchers at Northwestern University and the APS proposed a specialized compression algorithm that preserves an absolute error bound on the square root of the intensities \cite{huangFastDigitalLossy2021}, but without an encoding stage featured in most modern lossy compressors \cite{liangSZ3ModularFramework2023}.

\subsubsection{Compression Requirements}
Requirements for compression vary significantly by facility, beamline, and the techniques used at each beamline.  We summarize the state of the art below.

\paragraph{Analysis Metrics and Quality}
For femtosecond crystallography the electron density reconstruction is the key structure to preserve \cite{underwoodROIBINSZFastSciencePreserving2023}.  The electron density reconstruction is heavily influenced by the detection of Bragg spots, which are small regions of high intensity, located with tools such as peak-detector-v3 from LCLS \cite{hadian-jaziDataReductionSerial2021}.

Other beamline techniques have proposals for quantities to preserve that need further study.
For small-angle X-ray scattering (SAXS) and wide-angle X-ray scattering (WAXS) a key feature to preserve is peak position in spectral results.
For high-energy diffraction microscopy (HEDM) a key feature to preserve is the pseudo-Voight fit around peaks in the image, which are identified by the observation of local maxima \cite{sharmaMarinerhemantMIDAS2024}.
For coherent surface scattering imaging (CSSI) a key metric to preserve is a Fourier ring correlation in the reconstructed image \cite{kohoFourierRingCorrelation2019}.
For tomography, the image quality of the reconstruction is key for the tomographic structure reconstruction and preserving quality for downstream applications such as segmentation \cite{nikitinTomocuPyEfficientGPUbased2023}.
Researchers at the APS have also begun a study of qualities of interest for X-ray photon correlation spectroscopy (XPCS) that could be used to evaluate compression quality.

\paragraph{Performance Requirements}
LCLS has started benchmarking the compression performance of  SAXS/WAXS images from the LCLS-II data acquisition system.  Measurements indicate that LCLS would need close to 1,000 64-core nodes to reduce the data volume with SZ3, which LCLS feels is too large to be maintainable by the facility's small team.  In principle this performance could be increased by batching together the data from several LCLS shots before giving it to the compression, but this coupling of shots data would require significant changes to downstream analysis software that distributes different shots to CPU cores.  To alleviate this performance bottleneck, LCLS has begun researching running these algorithms on GPUs, where LCLS hopes to be able to process $\sim$50 GB/s per GPU, requiring only $\sim$20 GPUs for the largest 1 TB/s detectors.  Keeping as much of the processing of the data on the GPU as possible is critical to maintaining performance. Additionally, LCLS will use direct memory access (DMA) to send the data directly from the detector FPGAs to the GPU using GPUDirect over PCI Express, compress the data, and then store the data on the WEKA file system using GPUDirect Storage.  The only high-rate part of the data acquisition path LCLS anticipates needing the CPU for is a software-trigger decision where small data results (e.g., Bragg peaks) are communicated to the trigger-decision machines.
Similar efforts are underway at the APS but focus on XPCS, HEDM, and CSSI experiment types.

The need to store this data places requirements on compression ratios as well.
While entropy can vary per experiment, it is typically sufficiently low that lossless compression yields a compression ratio of only 2.
As a result, lossy compression is required.
Researchers at LCLS estimate that a compression ratio of at least 10 is required, which is possible using methods such as ROIBinSZ \cite{underwoodROIBINSZFastSciencePreserving2023}.  Similar compression ratios need to be reached for other beamlines and techniques at LCLS.

\paragraph{Sustainability}
Beamline scientists estimate that data needs to be retained for $\sim$10 years for most datasets or the lifetime of the beamline, whichever is longer. Additionally, some older datasets may be needed for calibration and comparison of older systems with newer systems.  As beamlines continue to become more advanced, however, the quality of datasets produced by these facilities continues to improve, which limits the usefulness of most older datasets.

\paragraph{Software Management}
Software management strategies can vary substantially by facility.
At the APS, software is largely either manually installed by beamline users/administrators or uses prebuilt Anaconda packages.
At SLAC, software was previously installed using Anaconda but now is being transitioned to Spack for its better handling of GPU-based libraries and applications compared with Anaconda.
Both facilities have a low-level component that is implemented in C/C++, with a higher-level user-facing component in Python, which is sometimes controlled via a graphical user interface.

\paragraph{Special Needs: Uncorrelated Dimensions, Small Buffer Compression, Hardware-Portable Decompression}
Light source data applications typically need three special capabilities that are not necessarily needed by other applications.

\textbf{Uncorrelated Dimensions} Like climate codes, light sources express their data with dimensions that may not necessarily be correlated.  However, unlike climate where the lack of correlation may be related to insufficient resolution, in light sources this can be caused by a dimension representing a non-contiguous quantity such as the panel id of an area detector that is assembled from many independent panels.

\textbf{Small Buffer Compression}  Beamlines produce many small images that need to be compressed.  This situatioin serves as a performance challenge for compressors that need large quantities of data to be presented simultaneously in order to achieve high compression ratios and high throughput.  This challenge is especially pronounced on the GPU where large quantities of data are needed to hide data movement costs.
Meeting this challenge may require capabilities such as GPUDirect Storage (or similar mechanisms from other accelerator vendors) to avoid unnecessary copies from CPU to GPU, optimized Huffman encoding mechanisms to achieve high compression ratios while retaining high performance per beamline, and improved concurrency in compression codes.

\textbf{Hardware-Portable Decompression} While devices on the beamline may feature state-of-the-art hardware to sustain the throughput from the detector, devices farther away from the detector where analysis is performed after experiments may feature less advanced hardware and thus may need the ability to decompress on different hardware from that originally used for compression.

\paragraph{Expected Changes}

As beamlines continue to advance, the data rates and resolution of detectors are expected to continue to increase, further stressing the computing and I/O subsystems of these facilities.
Beamlines are considering adopting increasingly sophisticated processes for data processing and compression, such as FPGAs, to keep up with these increasing data rates.

\subsection{Molecular Dynamics Simulations}
Molecular dynamics (MD) simulations examine the movement of particles within physical space to uncover the system’s dynamic progression based on particle interactions. These simulations have become a crucial research tool across numerous scientific fields, including physics, biology, and materials science. 
In biophysics and structural biology, MD simulations are widely used to investigate the behavior of macromolecules such as proteins and nucleic acids, facilitating the interpretation of biophysical experimental data and the modeling of molecular interactions~\cite{abraham_gromacs:_2015, noauthor_mddb_nodate}. 
In materials science, MD simulations enable researchers to model and predict the structural, thermal, and mechanical properties of materials at the atomic level. This capability helps in understanding phenomena such as material deformation, fracture mechanics, and phase transitions, providing insights that are often inaccessible through direct experimental observation~\cite{thompson_lammps_2022}.

\subsubsection{Motivations for Compression}
MD simulations have applications in numerous domains, where the generated data typically consists of particle coordinates as a function of time. 
The size of uncompressed binary coordinate trajectory files depends on the system size and simulation settings but is ordinarily tens to hundreds of gigabytes, emphasizing the need for efficient lossy compression algorithms---system sizes of biomolecular MD simulations commonly range from 100,000 atoms to a few million atoms, while the simulation time scales are often tens of nanoseconds to tens of microseconds, with trajectory frames writing intervals often in the range of 1 frame per 10 picoseconds to 1 frame per nanosecond~\cite{spangberg_trajectory_2011, lundborg_efficient_2014}. 

\subsubsection{Current Uses of Compression}
Some lossy compressors have been integrated in MD simulation packages. For example, the GROMACS~\cite{abraham_gromacs:_2015} package compresses coordinate trajectories using two different, but related, lossy methods. The XTC~\cite{xtc} file format was introduced in GROMACS 25 years ago and has been adopted by many projects needing efficient storage. The format uses external data representation routines for metadata portability between architectures, while the actual compressed data is a binary format that makes extensive use of correlations between subsequent atoms in water for efficient bit compression. The format does not support storage of either velocities or forces. The compression is very efficient, often achieving a compression ratio of 3--3.5, with a precision of 0.001 nm (corresponding to a maximum absolute error of 0.0005 nm), for water-rich systems. 
In order to employ a more modern file format and improved compression alternatives, roughly 10 years ago the TNG~\cite{lundborg_efficient_2014} file format was introduced in GROMACS. The TNG compression routines~\cite{spangberg_trajectory_2011} were based on XTC compression, with added temporal (multiframe) compression alternatives. When writing frames at short intervals, temporal compression with TNG can be much more efficient than XTC compression; however, when coordinates are written less often (>1 ps between frames), the gain is not substantial \cite{spangberg_trajectory_2011}. The TNG compression speed can be significantly slower than that of XTC, because of the Burrows-–Wheeler-–Lempel-–Ziv–-Huffman algorithm~\cite{burrows_block-sorting_1994,ziv_universal_1977,huffman_method_1952,bentley_locally_1986}. The TNG file format has not gained much traction in the general MD simulation community, partially because of its not being supported in some packages and also because the established XTC file format is good enough for most purposes.

\subsubsection{Compression Requirements}\label{sec:needs_md}

The main goal of using compression in MD simulation is to reduce the data storage needs without compromising simulation speed or increasing code complexity.
Important requirements are user-specified absolute error limits (possibly also relative error limits) and that the order of atoms and coordinates are recovered upon decompression. It is desirable that the compression/decompression is not significantly slower than when using XTC. Since the time spent writing a trajectory is negligible compared with the rest of an MD simulation (if not writing frames extremely often), whereas analyses and visualizations of a single trajectory may be done over and over, it is more important that decompression be fast.

\paragraph{Analysis Metrics and Quality}
Evaluating the quality of lossy compression for MD data involves two key aspects. First, it is crucial to preserve essential structural features such as chemical bonds, hydrogen bonds, and angle restrictions. This can be achieved by setting appropriate error bounds during compression, ensuring that deviations in bond lengths (e.g., within 0.1 Å), hydrogen bond distances (e.g., within 0.35 Å), and angular constraints remain within acceptable limits. 
Second, maintaining the consistent sequence of particles across all frames (snapshots) is essential for tracking individual particles over time. This continuity is necessary for many analyses, such as studying particle trajectories, diffusion processes, and conformational changes. 

\paragraph{Performance Requirements}
Adding compression to MD simulations requires minimal computational overhead to maintain simulation speed. It must support high I/O throughput to efficiently handle large data volumes and scale effectively on HPC systems through parallel processing. 
Moreover, since single-frame analysis is heavily used, the decompression of any individual frames must be fast enough to avoid slowing down the analysis workflow. 
The existing MD compressor, XTC, delivers satisfactory performance and should serve as a baseline for future compressor development. 

\paragraph{Sustainability}
MD simulation data usually has a retention requirement of 10 years or longer.
Long-term data retention supports ongoing and future studies by providing a valuable resource for reanalysis with new techniques or for exploring different research questions. One example is the the MDDB (Molecular Dynamics Data Bank) project~\cite{noauthor_mddb_nodate}, which aims to create the first unified database for MD simulations, providing a platform for scientists to share, access, and build on one another’s work, thereby accelerating discovery and innovation in the field.

\paragraph{Integrations and Installations}
When the MDDB project~\cite{noauthor_mddb_nodate}, funded by the European Union, started, it was decided that a modern and extensible file format for storing MD simulation trajectories was important in order to efficiently store files for database access and to help the community implement support for this format in a wide range of codes.
The file format should ideally support compression at least as efficient as XTC and TNG compression methods, with multiframe compression. It was desired to avoid having to write, and support, new file format libraries and APIs and preferably to use a specification that would be widely accepted. The decision was to use the H5MD~\cite{de_buyl_h5md_2014} specification, designed to store molecular simulation data in the HDF5~\cite{noauthor_hdf5_nodate} file format. As such, one fundamental requirement is that the compression algorithms be available as HDF5 compression filters.
Additionally, including the binary executables for compression is essential for testing and debugging purposes.

\paragraph{Special Needs}
In MD simulations, data is represented as a collection of discrete particles rather than a dense grid~\cite{abraham_gromacs_2024}. Each particle corresponds to an atom or molecule and carries attributes such as position, velocity, and force, evolving over time according to physical laws. Unlike grid-based methods, which store values at fixed spatial points, MD tracks individual particles in continuous space, making it well suited for capturing atomic-scale interactions and dynamics. This particle-based format results in sparse, high-dimensional data, where compression strategies must account for both spatial correlations and temporal evolution to effectively reduce storage while preserving essential physical properties.

\paragraph{Expected Changes}
In this workshop, the focus was on biomolecular MD simulations from the perspective of the GROMACS~\cite{abraham_gromacs:_2015,pall_heterogeneous_2020,abraham_gromacs_2024} software package and for storing MD simulation trajectories in databases, related to the MDDB project~\cite{noauthor_mddb_nodate}. 
A collaboration exists between the GROMACS and the SZ3 developers to improve support for biomolecular MD trajectories, based on the lossy MDZ framework~\cite{zhao_mdz_2022}. There are plans to implement XTC compression in the SZ3 compression framework to efficiently compress single frames, with more advanced compression routines used for multiframe compression.
Future needs and changes are expected to be similar for other MD simulation packages, such as AMBER~\cite{salomon-ferrer_overview_2013}, CHARMM~\cite{brooks_charmm:_2009}, LAMMPS~\cite{thompson_lammps_2022}, and NAMD~\cite{phillips_scalable_2020}.

\subsection{Quantum Circuit Simulation}

Quantum computing simulation is essential for advancing quantum computing, a rapidly growing field at the intersection of physics and computer science. Quantum physics enables the design of devices that have the potential to solve problems infeasible for classical computers. To develop and verify these technologies, quantum circuit simulators play a crucial role by allowing researchers to test quantum devices and evaluate quantum algorithms without requiring physical quantum hardware. These simulations help researchers optimize existing algorithms and explore new approaches. The goal of these simulations usually fits into two categories: finding a value of some observable and finding the probability of some set of states.



The challenges of simulation of a quantum computer come from the same source as its
potential advantage: as the system size grows, it requires an exponential amount of memory
to specify its state. This is also the source for the need for compression.


\subsubsection{Motivations for Compression}

The fundamental building block of a quantum computer is called a qubit. A
quantum system of $N$ qubits requires $2^N$ numbers to fully specify its state.
The simulation of a quantum system then involves applying a sequence of
operations, or ``gates'' to the state. There are two major types of quantum
circuit simulators: state vector simulators and tensor network simulators. 
The state vector approach stores the state as a vector of $2^N$ complex numbers 
and performs the gate applications as a sequence of matrix-vector multiplications.
The tensor network approach does not create the state vector and instead
combines the gates with each other as a sequence of tensor contractions to
obtain the final result.

State vector simulators face several significant bottlenecks that limit their scalability and efficiency in simulating large quantum circuits.
The most prominent bottleneck is the exponential growth of memory needed to
store the quantum state vector. 
This scaling quickly exhausts
available memory resources as the number of qubits increases, typically limiting
state vector simulations to about 45 qubits on existing supercomputers. The
computational cost of applying quantum gates also scales exponentially with the
number of qubits. Each gate operation involves multiplying the state vector by a
large matrix, resulting in intensive calculations that become prohibitively
time-consuming for high-depth quantum circuits.
When simulations are distributed across multiple nodes to handle quantum
circuits with a high number of qubits on supercomputers, the communication
between nodes becomes a significant bottleneck. Updating the state vector often
requires exchanging data between processes, which can dominate the simulation
time for certain types of circuits. While some parallelization is possible, the
inherently sequential nature of applying gates in a circuit limits the
efficiency gains from parallel computing resources, especially for deep
circuits.

The most significant bottleneck in tensor network simulations of quantum circuits is the storage of intermediate large tensors. As quantum circuits grow in size and complexity, the intermediate tensors produced during contraction can become extremely large (up to 16 PB with 50 qubits), potentially exceeding available memory.
When intermediate tensors do not fit in memory, they must be stored in distributed memory, which adds significant communication overhead and slows down the simulation process. It is done by using tensor slicing techniques. Another problem is finding the optimal contraction order for tensor networks. Since it is an NP-hard problem, suboptimal orders can lead to larger intermediate tensors and significantly increased memory requirements. As a result, tensor network simulators can simulate circuits with a relatively high number of qubits, typically up to 200 qubits (compared with the state-vector simulators), but with a short depth.

\subsubsection{Current Uses of Compression}

Physics research has a long history of representing a state as a product
of tensors, as opposed to a single dense tensor as in the state-vector approach.
Such representations are known as matrix product states \cite{MPS} or projected entangled pairs
 \cite{peps} and are constructed by using SVD or QR decompositions.
They are sometimes considered a
lossy-compressed representation of the original state but with a key
difference: they do not require decompression. One can apply gates to the
compressed representation directly. One can also calculate the final result, observable or probability, directly from this compressed representation.
However, these approaches require repetitive calculation of expensive
decompositions such as SVD and QR, which slows down the simulation.

One state-of-the-art exploration is applying lossy compression to a state-vector
simulator  
In ~\cite{fullstate-quantum-SC19,wu2018amplitudeawarelossycompressionquantum} 
this teachnique was used to increase the size of simulations by carefully balancing the trade-off between memory usage, computation time, and simulation fidelity. Quantum state vectors are represented by using complex numbers. The lossy compression method targeted these complex amplitude values. Pointwise relative error bounds were used to control the compression quality. The error bound determined the allowable difference between the original and compressed data values.

Another key technique is the use of an adaptive approach to error bounds. Initially, a tight error bound is applied. As memory constraints tighten, the error bound is relaxed incrementally, increasing the compression ratio. During simulation, the state vector is split into blocks, which are lossy compressed. Each block is decompressed when needed for computation and recompressed afterward.
Only two blocks ae decompressed at any given time to minimize memory usage.
By applying lossy compression, the memory required to store the quantum state vector is significantly reduced. For example, the memory required for simulating the 61-qubit Grover’s search algorithm was reduced from 32 exabytes to 768 terabytes.


In addition to state-vector-based quantum computing simulation, Shah et al.~\cite{shah-IPDPS23} developed a novel configurable compression framework tailored to the characteristics of quantum circuit tensor datasets generated by the state-of-the-art tensor network simulator QTensor. The framework incorporates a series of optimized preprocessing and postprocessing steps to enhance compression ratios with minimal performance overhead. Additionally, the study evaluated the impact of lossy decompression on quantum circuit simulation results, ensuring the fidelity of reconstructed data. To support GPU acceleration, the authors integrated their framework with cuSZ \cite{cuszplus} and cuSZx \cite{szx}, two leading GPU-based lossy compressors, offering configurable trade-offs between compression ratio and speed. Experimental results on an NVIDIA A100 GPU, using QTensor-generated tensors of varying sizes, demonstrated that the proposed strategies achieve nearly 10× higher compression ratios compared with cuSZ alone. When prioritizing throughput, the framework maintains compression speeds comparable to those of  cuSZx while achieving 3–-4× higher compression ratios. Moreover, decompressed tensors enable QTensor circuit simulations to produce final energy results within 1--5\% of the true energy value.

\subsubsection{Compression Requirements}


The compression requirements encompass two key aspects: analysis metrics/quality and performance expectations. Previous studies \cite{fullstate-quantum-SC19,shah-IPDPS23,Zhang:2024pub-qc-compression} have highlighted the critical need for lossy compression in quantum circuit simulation, examining user requirements and evaluating its impact on simulation performance and accuracy.


\paragraph{Analysis Metrics and Quality}

Evaluating the compression quality involves comparing the simulation result from the compressed simulation with the result of a lossless one. As mentioned in the introduction, there are two types of simulation results: an expectation value of an observable, which is a scalar real number, and a set of probabilities for quantum states, represented as a vector of real numbers between 0 and 1. These results can be compared with a reference value using metrics such as relative difference, L2 norm, or cosine similarity \cite{wu2018memoryefficientquantumcircuitsimulation,wu2018amplitudeawarelossycompressionquantum,Zhang:2024pub-qc-compression}. Additionally, some quantum circuits preserve specific observables as physical invariants, and their preservation can serve as an accuracy metric. A particularly important quality metric in quantum circuit simulation is fidelity \cite{fullstate-quantum-SC19}, which measures the similarity between the compressed and ideal quantum states. The fidelity metric provides a lower bound on simulation accuracy by estimating the cumulative effect of lossy compression across all quantum gates. It ensures that the reconstructed state maintains a high degree of similarity to the original, thus validating the effectiveness of the compression strategy.

Such comparisons are easy to make once the reference simulation outputs are obtained. For large-scale problems, however,  the reference results may be unable to be obtained; thus, calculating the simulation quality would be impossible.

\paragraph{Performance Requirements}

An effective compressor for quantum circuit simulation must balance compression ratio, throughput (or speed), and error to maximize performance. Compression algorithms are applied in an online fashion as part of the main simulation loop, making low compression/decompression overhead essential. The primary bottleneck in quantum circuit simulation is RAM usage, and applying compression can significantly expand the scale of simulations by reducing memory requirements.  

To be impactful, a compressor should achieve a compression ratio significantly above 2, ideally around 10$\times$ \cite{wu2018amplitudeawarelossycompressionquantum,fullstate-quantum-SC19,Zhang:2024pub-qc-compression}. The total memory required for storing the dataset is given by $16 \times 2^K$ bytes, where 16 represents the bytes needed for a double-precision complex number and $K$ is the number of qubits in the simulation. If the state vector is compressed by a factor of $N$, the simulation scale can be increased by $\log_2 N$, as demonstrated in Wu et al.'s work \cite{fullstate-quantum-SC19}.  

Equally important is the throughput of compression and decompression, as it directly impacts overall simulation time. Unlike traditional memory access, a compression-supported simulation must frequently decompress data for calculations and recompress it afterward. This overhead can be substantial, especially with computationally expensive compression algorithms. As shown in Wu et al.'s study \cite{fullstate-quantum-SC19}, even with a lightweight compressor design, compression overhead can account for up to 90\% of total simulation time in the worst case. Therefore, achieving high compression throughput is critical to minimizing performance degradation while enabling larger-scale quantum circuit simulations.

Modern simulations can be highly I/O-intensive, making GPUs a common choice for accelerating computations. In order to maintain efficiency, it is also preferable to perform compression on GPUs, leveraging their parallel processing capabilities to minimize data movement overhead and enhance overall performance.

\paragraph{Sustainability}
At the moment, the motivation for applying compression in quantum circuit simulation is mainly for the real-time RAM compression use case, so the data is not expected to be stored for a long time. Typical examples include the online compression of quantum state vectors in Wu et al.'s work \cite{fullstate-quantum-SC19} and online compression of quantum circuit tensor datasets in Shah et al.'s work \cite{shah-IPDPS23}. 

\paragraph{Installation and Integration}

The most common language in the area is Python, with widespread use of C++ bindings.
The input data is usually passed as a GPU device pointer to a buffer of complex numbers
with an option to configure the precision.
Running compression in parallel with other tasks in the simulation may benefit the performance.
Installation through pip or Anaconda package managers is preferred.

\paragraph{Special Needs: High-Dimensional Data, Complex Datatype, Partial Decompression}

\textbf{High Dimensionality} The tensors in quantum circuit simulation represent probability amplitudes for
some quantum events. A typical tensor has up to 30 dimensions and contains small
complex values. Each tensor dimension is small, usually equal to 2. Because
of the high dimensionality (e.g., ndims > 128), tensors may exhibit periodic patterns with a period of
powers of 2. Exploring ways to leverage the unique characteristics of quantum circuit data for enhancing lossy compression design presents a promising research direction.

\textbf{Leveraging Sparsity Patterns} 
Regarding the state-vector-based quantum circuit simulation, Wu et al.~\cite{fullstate-quantum-SC19} developed the lossy compressor by making full use of the data sparcity in two aspects. (1) \textit{Lossless Compression for Sparse Data}: At the beginning of the simulation, when most values are zero, lossless compression (such as Zstd) is effective in reducing memory usage while preserving full data fidelity. (2) \textit{Bit-Plane Truncation}: Later in the simulation, as the data becomes more complex, an error-bounded lossy compressor is applied, which includes truncating insignificant bit-planes based on a user-defined relative error bound.
As for the tensor-based quantum circuit simulation \cite{shah-IPDPS23}, the amplitudes are multiplied with each other during the course of simulation, so very small values can be common. In general, the values scale as $2^{-N/2}$ with problem size. The properties of tensors  may change between different types of quantum circuits. 
For example, some circuits result in concentration of probability amplitudes, which result in more sparse tensors. To address this sparsity feature, the developed compressor \cite{shah-IPDPS23} employs two key techniques. (1) \textit{Thresholding Small Values}: A threshold filter is applied to tensor values, setting those below a certain threshold to zero. This increases data similarity, which in turn improves the efficiency of quantization-based lossy compression. (2) \textit{Threshold+Grouping Method}: Instead of storing all tensor values, this method separates nonzero (significant) values into a ``significant value array,'' while a bitmap records their original positions. This reduces storage overhead by avoiding the need to store and process many zero values, further boosting compression efficiency.



\paragraph{Expected Needs}
The time overhead of compression and decompression can significantly slow down the overall simulation, since data in a compression-free simulation can be accessed and stored directly in memory without additional processing. To address this issue, homomorphic compression \cite{hoszp}, which enables numerical operations to be performed directly on compressed data, presents a promising approach. By reducing the need for frequent decompression and recompression, homomorphic compression could greatly enhance the efficiency of compression-supported quantum circuit simulations.

\subsection{Seismology}

Seismic imaging is a technology that creates high-fidelity Earth's subsurface images by analyzing the propagation and reflection of seismic waves.
In energy industries, companies such as Saudi Aramco utilize seismic imaging to optimize resource (e.g., oil) extraction while minimizing environmental impact~\cite{huang2023towards,malovichko2025towards}.
Seismic imaging is also essential in various domains~\cite{li2022research,daramola2024enhancing,doerr2008development}, such as assessing the stability of tunnels and bridges~\cite{kase2004seismic}, and even in planetary sciences~\cite{zhao2008seismic}, where it helps study the internal structures of celestial bodies such as the moon and Mars.
Given its wide-ranging applications, improving the efficiency and accuracy of seismic imaging is critical for both scientific advancements and industrial applications.

\subsubsection{Motivations for Compression}
In this section we provide two examples that have big data issues from computational seismology where the goal is to use 3D numerical wave simulations to image Earth's interior, with an emphasis on global-scale adjoint tomography \cite{tromp2005seismic}, a full-waveform inversion (FWI) technique~\cite{WoDz84,Lailly1983,Tarantola84a}, and reverse time migration (RTM)~\cite{baysal1983reverse}, a high-resolution seismic imaging method that reconstructs subsurface reflectors by back-propagating recorded wavefields using a given velocity model.

\begin{figure}[!htbp]
  \begin{center}
  \includegraphics[width=0.8\columnwidth]{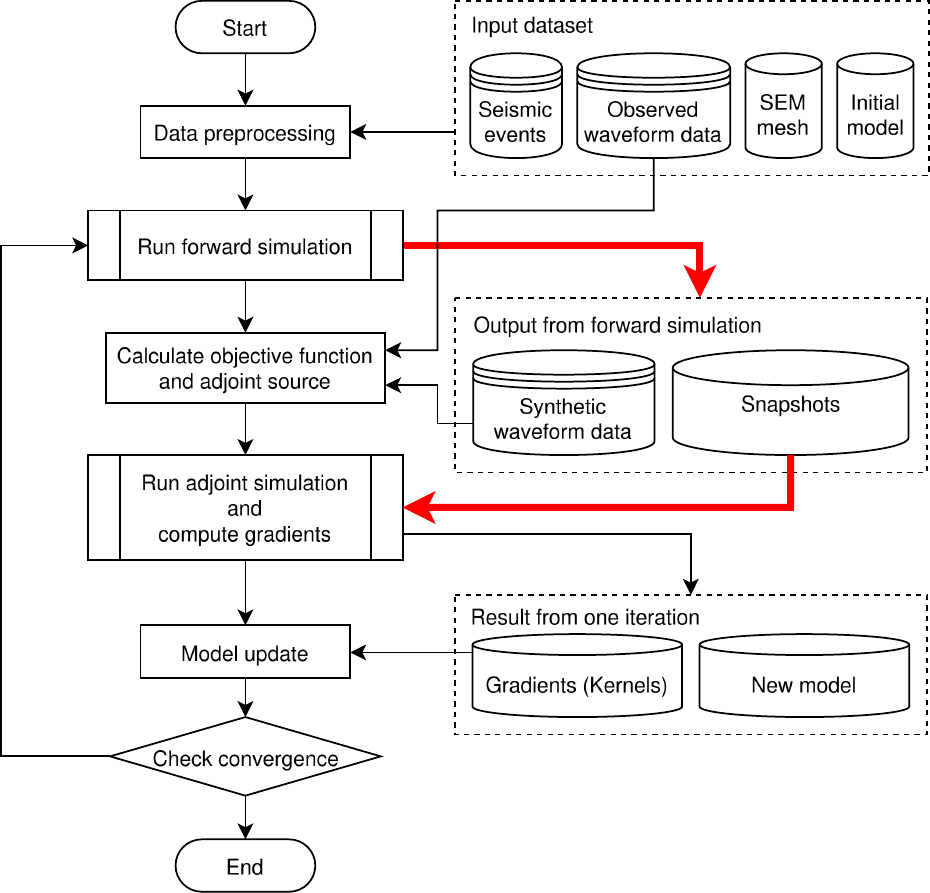}
  \end{center}
  \vspace*{-1em}
  \caption{Diagram of FWI workflow. In a iteration loop, there are two I/O peak timings indicated by the red arrows. The first peak occurs when all the simultaneous forward simulations store their snapshots on physical storage. The second peak occurs when the simultaneous adjoint simulations read those snapshots to recreate the wave fields.}
  \label{fig:seismo_flow}
\end{figure}

Adjoint tomography
integrates the full physics of wave propagation into seismic imaging by computing synthetic seismograms and data sensitivity kernels, also known as adjoint or Fr\'echet kernels, using spectral-element solvers such as SPECFEM3D\_GLOBE~\cite{Koma02a,Koma02b}.
The adjoint method involves a \textit{forward wavefield}, generated by a seismic source and recorded at receiver locations, and an \textit{adjoint wavefield}, which backpropagates waveform misfits to refine model parameters~\cite{tromp2005seismic}.
FWI is an iterative optimization process that minimizes the difference between observed and synthetic waveforms to improve subsurface models, requiring repeated forward and adjoint simulations.
However, large-scale FWI workflows generate massive data volumes, since wavefield snapshots must be stored and retried during adjoint computations.
Because of memory constraints, these snapshots are written to disk, creating significant I/O bottlenecks when reading and writing volumetric data, especially when processing multiple seismic events concurrently.
As shown in Figure~\ref{fig:seismo_flow}, these I/O peaks occur at red arrows in the workflow.
At these timings, all the MPI processes perform write/read operations on volumetric data arrays simultaneously.
Moreover, when multiple seismic events are calculated simultaneously, the consumption of I/O bandwidth increases proportionately to the number of simultaneous runs.
Consequently, data compression is essential to mitigate these challenges, reducing storage requirements and improving computational efficiency in large-scale seismic inversions.

\begin{figure}[!htbp]
  \begin{center}
  \includegraphics[width=0.8\columnwidth]{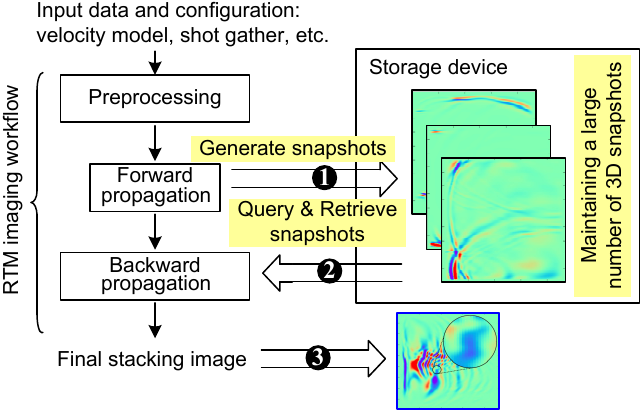}
  \end{center}
  \vspace*{-1em}
  \caption{Illustration of the big data issue in preserving forward propagation wave snapshots during runtime of reverse time migration (RTM) execution.}
  \label{fig:rtm-illus}
\end{figure}

RTM reconstructs high-resolution subsurface images by backpropagating recorded seismic waves, relying on an accurate velocity model. Similar to adjoint tomography, RTM execution demands significant storage and bandwidth due to the large volume of wavefield snapshots involved. Figure~\ref{fig:rtm-illus} illustrates the workflow of an industrial-scale parallel RTM implementation, representing a practical seismic imaging process. At the start of execution, RTM is initiated by using input parameters, including \textit{problem size}, \textit{initial background data file} (i.e., velocity data), \textit{total number of snapshots}, and the \textit{interval} $K$ at which snapshots are saved for subsequent backpropagation analysis. During forward propagation, the source wavefield generates snapshots at each time step. Instead of storing all snapshots, however, only a subset (e.g., every $K$ time steps) is retained for later analysis, while the rest are discarded (step 1 in Figure~\ref{fig:rtm-illus}). Traditionally, these selected snapshots are either kept in memory---if resources permit---or temporarily written to external storage. Once forward propagation is complete, the backward propagation of the receiver wavefield begins, requiring access to the previously stored snapshots for subsurface imaging (step 2 in Figure~\ref{fig:rtm-illus}). 
Given the massive data volume, efficient data reduction techniques, such as compression, are essential to minimize storage overhead and improve I/O efficiency.
After the backward propagation phase, the final subsurface image is generated through a stacking process (step 3 in Figure~\ref{fig:rtm-illus}).

\subsubsection{Current Uses of Compression}
Compression for seismic imaging to reduce I/O burden, memory footprint, and storage overhead has been widely studied in the past decade~\cite{zand2019compressed,boehm2016wavefield,huang2023towards,barbosa2023reverse,soares2024study,maltempi2024combining}.
For adjoint tomography, Boehm et al.~\cite{boehm2016wavefield} addressed memory and I/O challenges in FWI by introducing a lossy compression approach for storing wavefield snapshots, which are crucial for adjoint-based seismic imaging.
The method combines temporal compression using cubic spline interpolation~\cite{sz3}, spatial compression with adaptive floating-point precision, and shadow zone detection to eliminate redundant data.
Integrated into finite-element wave propagation codes, this technique reduces storage up to three orders of magnitude while maintaining accuracy in sensitivity kernels.
With minimal computational overhead (2--10\%), it offers a practical alternative to checkpointing, significantly improving I/O efficiency in large-scale seismic inversions.
On the other hand, RTM requires a high-speed, in situ parallel lossy compression solution to manage its massive data footprint and alleviate I/O bottlenecks.
For example, Huang et al.~\cite{huang2023towards} designed HyZ, a hybrid OpenMP-based parallel lossy compressor integrating blockwise regression~\cite{Xin-bigdata18,sz-auto} and ultrafast bit-manipulation compression (SZx)~\cite{szx} to achieve both high compression ratios and minimal computational overhead.
By compressing forward propagation snapshots before storage and decompressing them on retrieval, this approach significantly reduces RTM’s I/O costs and memory demands while maintaining high data fidelity. 
Results show that HyZ improves overall RTM execution by 6.29--6.60×, outperforming other state-of-the-art compressors in both speed and compression efficiency.
Additionally, GPU acceleration has emerged as a promising approach for enhancing RTM execution, leveraging massive parallelism to handle large-scale seismic computations efficiently. This makes pure-GPU ultrafast lossy compressors, such as cuSZp~\cite{huang2023cuszp,cuszp2} and FZ-GPU~\cite{fzgpu}, ideal solutions in RTM workflows. By performing compression and decompression entirely on the GPU, these methods minimize data movement and maximize throughput, further accelerating seismic imaging.



\subsubsection{Compression Requirements}

Seismic imaging applications impose three key compression requirements: maintaining data quality in velocity models and high-resolution stack images, achieving high compression and decompression throughput, and obtaining an effective compression ratio. This section details each of these aspects~\cite{huang2023towards,barbosa2023reverse}.

\paragraph{Analysis and Quality Requirements}

As demonstrated in previous studies~\cite{boehm2016wavefield,huang2023towards}, seismic imaging applications impose strict requirements on reconstructed data quality, since any degradation can impact subsurface interpretation and decision-making. While intermediate data quality, such as individual wavefield snapshots in RTM, can tolerate some level of loss, the velocity model and final stacking image must maintain high fidelity to ensure accurate seismic imaging results. Unlike general image or video compression, where metrics such as SSIM~\cite{wang2004image} and PSNR~\cite{hore2010image} serve as standard quality indicators, seismic applications rely on domain expert analysis, where even small visual distortions can be unacceptable. The reason is that seismic imaging is highly sensitive to artifacts, phase shifts, and amplitude distortions, which may lead to misinterpretations of geological structures. Therefore, effective compression strategies must minimize loss in critical regions.
Additionally, a moderate compression ratio (e.g., 5× or higher for RTM) is sufficient as long as it efficiently reduces disk I/O burdens and enables data movement within memory, ensuring overall computational efficiency.

\paragraph{Performance Requirements}
Throughput is critical, since compression in seismic imaging is always performed in situ, meaning it must operate with minimal overhead to avoid slowing down the workflow. For GPU-based seismic imaging, compression must be as fast as possible to fully utilize the massive parallelism of modern accelerators. This makes ultrafast GPU compressors, particularly single-kernel designs such as cuSZp~\cite{huang2023cuszp,cuszp2}, highly effective for real-time RTM execution. For example, Saudi Aramco mandates that compression and decompression throughput for GPU RTM must be performed entirely on the GPU, with a minimum speed of 100 GB/s on an NVIDIA V100 GPU, ensuring that storage and memory constraints do not become bottlenecks in large-scale seismic imaging pipelines.




\paragraph{Sustainability}
Compressed data in seismic imaging is typically short-lived~\cite{boehm2016wavefield,barbosa2023reverse}, since it is used primarily for real-time execution, whether for checkpoint-restart~\cite{maltempi2024combining} or I/O reduction~\cite{huang2023towards}. 
For example, intermediate snapshots in RTM are needed only for backpropagation; once the final high-resolution stacking image is generated, storing these snapshots becomes unnecessary.


\paragraph{Integration and Installation}
Seismic imaging projects are predominantly C/C++-based, compiled with \verb|-O3| optimizations, and designed for high-performance parallel execution, where intermediate data is typically handled as pointer arrays. For CPU-based parallel execution, a single API should be provided with OpenMP support, ensuring efficient multithreading and achieving a throughput of at least 10 GB/s. For GPUs, performance demands are even higher, favoring a pure-GPU, single-kernel design with a throughput exceeding 200 GB/s to fully utilize modern accelerators. In order to facilitate seamless integration and deployment, the compression library should be well structured, supporting both static and dynamic linking, and ideally CMake-compatible for ease of use in large-scale seismic workflows.

\paragraph{Special Needs}
Seismic imaging data presents unique characteristics that require specialized compression strategies to optimize both compression ratio and throughput. 
Since most seismic datasets exhibit wavelike patterns~\cite{boehm2016wavefield,sz3,huang2023towards}, fast and fine-tuned interpolation is essential to enhance compression efficiency. 
Additionally, in applications such as high-resolution RTM~\cite{huang2023cuszp}, the data behaves as a time series, where early time steps tend to be sparse with a large value range, making them highly compressible, whereas later time steps become denser with a lower value range, making compression more challenging. A dynamic error bound is necessary to adaptively balance compression quality and efficiency. 
From a performance perspective, throughput must be maximized, particularly in I/O-heavy stages, ensuring that data movement does not become a bottleneck. 
For GPU-integrated compressors~\cite{huang2023cuszp}, the compression and decompression kernels must maintain a minimal memory footprint, avoiding techniques such as register spilling~\cite{chaitin1982register}, which can introduce unnecessary global memory overhead and negatively affect original execution effectiveness.

\paragraph{Expected Future Changes}
Seismic imaging compression will continue evolving to enhance adaptability, GPU efficiency, and HPC integration for large-scale workflows~\cite{huang2023cuszp,huang2023towards}.
(1) Adaptive Compression: Dynamic error bounds will optimize accuracy and efficiency across different imaging stages.
(2) GPU-Centric Designs: Ultrafast, single-kernel GPU implementations will minimize CPU-GPU transfers and push throughput beyond 500 GB/s on latest GPU variants (e.g., NVIDIA H100).
(3) Seamless HPC Integration: Closer integration with parallel I/O libraries (e.g., HDF5) will further reduce storage and I/O bottlenecks.


\subsection{System Logs}

Large-scale parallel and distributed applications generate enormous volumes of logging and monitoring data that must be aggregated and interpreted to understand the progress and performance of applications. One common example is the use of scientific workflows to orchestrate various scientific applications. Parsl~\cite{babuju19parsl}, Ray~\cite{moritz18ray}, TaskVine~\cite{delgado23taskvine}, and Globus Compute~\cite{chard20funcx} are examples of systems that coordinate execution of many different tasks on parallel and distributed infrastructure. 

\subsubsection{Motivations for Compression}

One challenge with task-based parallel and distributed applications, such as workflows and function-as-a-service platforms, is the need to monitor execution performance of various tasks on parallel and distributed systems. Monitoring information is used both interactively in real time and after execution to investigate how an application performed, detect anomalies and guide scheduling decisions. Collecting sufficiently fine-grained performance information, particularly for tasks that run for short periods, can place significant overhead on the monitoring infrastructure employed by the workflow software. 
However, the monitoring information is not stochastic. It follows various patterns, may remain static for long periods of time, and need not be high precision for many use cases.  

The monitoring system is generally implemented as one or more processes deployed alongside \textit{workers} deployed on provisioned nodes. These processes use both application and 
system monitoring information (e.g., via Python's psutil library) to measure resource use. The processes return monitoring information to the central workflow system via different methods. For example, Parsl enables this information to be returned by using a UDP-based protocol sent to a waiting monitoring hub, over the control channel used to manage workers, or via external mechanisms such as Apache Kafka. The Parsl workflow may use this information to make scheduling decisions; however, more commonly it is used by users to introspect workflow performance via queries to a database or through the Parsl monitoring web interface while the workflow is running or after completion. 

Providing rich monitoring information can represent a significant amount of data, as monitoring information is captured at a subsecond granularity from each worker in the system. Workflows running on a supercomputer may therefore have hundreds of thousands of workers concurrently capturing resource use.
The implications of a significant monitoring burden are that
workflows may exhibit poor performance when monitoring is 
enabled. For example, Parsl workflows have incurred up to an order-of-magnitude throughout degradation when monitoring is enabled~\cite{kerney24parsl}. Further, loss of monitoring information or delayed transmission can affect scheduling decisions, leading to reduced workflow performance. 

\subsubsection{Current Uses of Compression}

Compression has been widely used to reduce storage 
space used by system logs~\cite{li24logshrink,yao22log,balakrishnan06lossless, chang24turbolog}. 
These methods exploit characteristics of 
log data to improve compression rates for long-term lossless compression.
The methods have been designed primarily to support system logs, with 
little prior work focusing on task-based logging and real-time monitoring. We are not aware of prior work applying lossy compression
to task monitoring and logging data. 

\subsubsection{Compression Requirements}
This use case represents an opportunity to exploit lossy 
compression to reduce the monitoring data transfer and storage burden.
Analysis of monitoring overheads in Parsl showed up to an order of magnitude reduction in throughput when monitoring is enabled~\cite{kerney24parsl}.
In an interactive monitoring setting, for example, for scheduling or user management of workflows, the compression must be performed in a high-throughput mode in which monitoring packets, or batches of monitoring packets, sent by each node to the central monitoring hub must be compressed.  For longer-term persistent storage the monitoring data
can be captured and stored locally with less stringent performance requirements. 


We focus on two primary use cases: (1) communicating real-time monitoring data for use in scheduling decisions and (2) persisting logging data for post facto 
analysis. 

\paragraph{Analysis and Quality Requirements}
Key quality requirements include accuracy and fidelity, allowing acceptable error thresholds and ensuring timely delivery with minimal latency. Additionally, compression must preserve important trends and patterns while being computationally efficient and scalable to handle the monitoring data's throughput without introducing significant overhead. For example, when used to analyze energy efficiency~\cite{kamatar2024greenfaas}, it is critical that load for each phase of a task be captured and communicated to provide accurate measurements. Similarly, when used for scheduling~\cite{kumar21continuum}, it is important that peaks be tracked because these directly correspond to the workload that can be accommodated without slowdown.  

\paragraph{Performance Requirements}
The primary goal of the real-time monitoring use case is to provide 
monitoring data as quickly as possible to the scheduling component.
In this case, reducing the load on the communication channels (e.g., shared file systems or HPC networks) is important. Such logging workloads can
place significant strain on shared file systems that results in 
large performance desegregation for the workflow system and other workload on the system. 

The second use case, persisting for post facto analysis,  primarily requires that data be stored efficiently and accurately with minimal impact on the workflow itself. This use case is therefore more tolerant of compression delays, particularly if it uses fewer resources and results in higher-quality data. 

In both cases, the specific requirements are application specific and
would change depending on needs. For scheduling, we ideally need sufficient quality data to inform the scheduling algorithm.  Analysis 
of different compression ratios with various scheduling algorithms is needed to determine levels of compression. Archival storage is dependent on the questions users would like to ask of the data after execution. In most cases these questions are looking for anomalies (e.g., "why did my workflow fail? which task used too much memory?") or for general utilization (e.g., "how well utilized was my resource?"). In both cases, it is important to quantify the cost of compression in terms of resources used to compress/decompress data and consider the trade-offs with respect to the communication and storage cost of the monitoring data. 

\paragraph{Sustainability}
Monitoring data used for scheduling need not
be stored for long periods of time.
While data may be retained to train predictive models and improve scheduling algorithms, the data used for scheduling is typically used only for in near-real time to aid scheduling and placement decisions. Aggregate information may be kept for longer periods of time (e.g., average runtime or resource use of a particular task type).
Monitoring data used for post hoc analysis requires that compressed data
be stored for longer periods of time, and data is rarely
decompressed for analysis. Data is stored to provide a level of provenance for executions, for subsequent analysis of performance, to investigate errors that were identified after execution, or to train models used for 
online scheduling and prediction. 

\paragraph{Integration and Installations}
The compression capabilities must be accessible to 
the monitoring, logging, scheduling, and analysis
components of task-based systems.  Thus, data should be delivered as a library that is installable and can be integrated directly in the various components of the task-based system, such as the worker components deployed on HPC nodes as well as the management components deployed on HPC login nodes, users' PCs, and cloud-hosted nodes.  In many cases these systems are written in Python.

\paragraph{Special Needs}
While the majority of monitoring data is represented
as floating-point numbers, logging data may also include other information such as text describing lifecycle and errors.  Online use of monitoring data may look at metrics over a sliding window to make assessments of current conditions (e.g., resource availability). 

\paragraph{Expected Needs}
As systems become more heterogeneous and applications more diverse, we see the types of monitoring data changing. For example, increasingly monitoring information is derived from GPUs, and applications include machine learning models that are used to make decisions. Similarly, the types of scheduling algorithms and post hoc analysis are increasingly leveraging machine learning methods to make sense of large monitoring data and make online decisions. Thus, the monitoring data would benefit from being accessible to machine learning models, and the ultimate measure of compression utility is the accuracy of the models in which the data is used. 


\section{Compression Technologies}

In this section we highlight the principles, error controls, hardware support, unique features, history, and impacts of the leading compressor technologies.
We begin with more established compressor frameworks including SZ, ZFP, and MGARD developed during the U.S. DOE Exascale Computing Project.
After that, we highlight some upcoming compressors and frameworks including LC, SPERR, DCTZ, and TEZip.
We conclude with a discussion of LibPressio, which is not a compressor itself but provides a common abstraction atop the various compressors.

\begin{table*}[]
    \centering
    \begin{tabularx}{\textwidth}{llllX}
    \toprule
         Compressor & Principle & Error Control & Hardware & Unique Feature \\
    \midrule
         SZ & prediction & various & various & flexibility  \\
         \zfp & transform & various & various & alternative floating-point format  \\
         MGARD & finite-element/wavelet & various & various & progressive, extensive error controls  \\
         LC & components and preprocesors & ABS,REL & CPU+GPU & byte-for-byte multi-hardware  \\
         SPERR & wavelet & ABS & multicore-CPU & progressive and multiresolution  \\
         DCTZ & transform & ABS & CPU & advanced quantization techniques  \\
         TEZip & recurrent CNN & ABS & various & time series and learning  \\
         LibPressio & abstraction & extensible & extensible & application focused  \\
    \bottomrule
    \end{tabularx}
    \caption{Overivew of compressors}
    \label{tab:compressor_summary}
\end{table*}

\subsection{SZ}\label{compressor:sz}
SZ (\url{https://szcompressor.org}) is a prediction-based error-bounded lossy compressor. In fact, it is not only an off-the-shelf general-purpose lossy compressor but also a composable framework allowing users to customize appropriate/effective compressors for specific applications or use cases. 

\subsubsection{Principles}\label{compressor:sz:principles}

The SZ family of compressors generally contain three critical steps: pointwise data prediction, quantization, and lossless integer encoding.

\textbf{Pointwise Data Prediction}. There are two strict constraints for the data prediction methods to be used in SZ. On the one hand, considering that the predicted data values must be identical between compression and decompression, the original raw data values cannot be directly used in the course of prediction (note that decompression stage has no original data information), That is, the prediction needs to be performed based on the decompressed/reconstructed data; otherwise, compression errors cannot be bounded as expected. On the other hand, the prediction policy should be able to go over every data point just one time, considering that the data values would be reconstructed one by one in the course of decompression. Data prediction is arguably the most critical step in the SZ compression pipeline because the more accurately the data is predicted, the more effective the succeeding compression steps will be.  
    
\textbf{Quantization}. Quantization divides a value range into consecutive non-overlapped intervals (i.e., quantization bins), which can transform the floating-value domain to an integer domain such that the succeeding compression operation would be very effective. The simplest (also mostly commonly used) quantization method is linear-scale quantization, where each quantization bin has fixed length, which is often used in general-purpose absolute error-bounded compression such as SZ1 \cite{sz17}, SZ2 \cite{Xin-bigdata18}, SZ3 \cite{sz3}, and cuSZp \cite{huang2023cuszp}.

\textbf{Integer Lossless Encoding}. After the quantization step, SZ adopts a series of lossless encoding operations on the integer values that were generated in the quantization step to significantly reduce the data size. The general lossless encoder adopted in SZ is a customized Huffman encoder \cite{sz17} followed by a dictionary encoder LZ77 \cite{lz77} (using Zstd \cite{zstd}), 

\subsubsection{Error Controls}
SZ supports different types of error control methods, including absolute error bound, value-rage based error bound, relative error bound, and peak signal to noise ratio (PSNR) \cite{z-checker}. Although the classical SZ framework (such as SZ2/SZ3) does not support fixed-ratio compression, the SZ team developed the Surrogate Error-bounded Compression Framework (SECRE) \cite{secre}, which provides support to enable fixed-ratio compression for different compressors including SZ. Specifically, SECRE allows one to accurately estimate the compression ratio by emulating the corresponding compressors' behavior/operations on sampled datasets, so that it can estimate the compression ratio based on a given error bound quickly and accurately. Many other emerging machine learning or statistical-analysis-based methods \cite{rahman2023feature,ArkaCluster23} also can be used for SZ's compression ratio estimation.

\subsubsection{Hardware Support}
The SZ team has developed different versions of SZ to adapt to diverse devices, including CPU, GPU, FPGA, and AI accelerators. For example, SZ1/SZ2/SZ3, SZ-auto \cite{sz-auto}, AE-SZ \cite{ae-sz}, Pastri-SZ \cite{pastri}, MDZ \cite{zhao_mdz_2022}, and CliZ \cite{cliz} were developed mainly for CPU architecture. FZ-GPU \cite{fzgpu}, cuSZ \cite{cuszplus}, and cuSZp \cite{huang2023cuszp} were developed for GPU devices (such as CUDA architecture) in particular. SZx \cite{szx} supports both CPUs and GPUs. WaveSZ \cite{wavesz} and VecSZ \cite{vecsz} were optimized for FPGA and single instruction, multiple data instruction sets, respectively. CereSZ \cite{ceresz} is a version that was developed/optimized for Cerebras CS-2 to address the specific needs of compression on AI accelerators.

Unlike other compressor frameworks,  the SZ framework of compressors tends to favor high performance on each platform rather than byte-for-byte compatibility between compressor implementations on different hardware platforms. However, this situation  is improving as a result of the FZ project.
To illustrate, we consider the case of CereSZ.
CereSZ is designed for the Cerebras CS-2 AI accelerator, which is based on the control flow architecture and does not have access to global memory. Each computing unit can access data only from a small local memory and its neighboring units. This restricted memory access presents new challenges, preventing large Huffman trees. As a result, CereSZ-2 features a fixed-sized Huffman tree not used on other platforms.

\subsubsection{Unique Features}
Unlike other traditional compressors, SZ3 is not only a compressor but a composable framework, which allows users to create diverse compressors by easily implementing/customizing specific methods in five different stages: data preprocessing (e.g., transforming raw data to log domain for pointwise relative-error-bounded compression \cite{liang2018efficient}), prediction (e.g., Lorenzo, linear regression \cite{Xin-bigdata18} and spline interpolation \cite{sz3}), quantizer (such as linear-scale quantization \cite{sz17}), encoder (such as Huffman encoding), and lossless compressor (such as Zstd \cite{zstd}).   

\subsubsection{History and Impact}

The SZ team has explored data prediction methods to adapt to diverse applications/use-cases, including Lorenzo predictor (used by SZ1 \cite{sz16,sz17}), linear regression (used by SZ2 \cite{Xin-bigdata18}), dynamic spline interpolation (used by SZ3 \cite{sz3,sz3modular}), autoencoder-based prediction (used by AE-SZ \cite{ae-sz}), scaled-pattern-based prediction (used by Pastri-SZ \cite{pastri}), wavelet-transform-based prediction (used by FAZ \cite{faz}), and climate-property-based prediction (used by CliZ \cite{cliz}).

The SZ team has also explored alternative quantization schemes.  A specific version \cite{9680367} of the SZ family allows the quantization bins to be of different lengths to adapt to diverse requirements on different value intervals/ranges. MDZ \cite{zhao_mdz_2022}---a customized version for MD simulations---supports vector quantization to adapt to clustering data patterns in MD datasets. 

Huffman encoding+zstd is the most common form of encoding adopted by many SZ family products such as SZ2 \cite{Xin-bigdata18}, SZ3 \cite{sz3}, cuSZ \cite{cuszplus}, MDZ \cite{zhao_mdz_2022}, QoZ \cite{qoz}, FAZ \cite{faz}, and HPEZ \cite{hpez-sigmod24}. Some variants, however, avoid the expensive cost of Huffman+zstd encoding on GPUs. FZGPU \cite{fzgpu} adopts a shuffle-based fixed-length encoding, and cuSZp \cite{huang2023cuszp, cuszp2} adopts a blockwise fixed-length encoding. 

The SZ family of compressors is an R\&D 100 award winner.

\subsection{ZFP}

\zfp (\url{https://zfp.io}) is primarily an in-memory compressed representation for multidimensional floating-point arrays that supports high-speed read and write random access at very fine granularity.

\subsubsection{Principles}

The \zfp backend is responsible for compressing and decompressing individual blocks via a pipeline of largely reversible steps.
\zfp compression takes a block of floating-point numbers and aligns them to a common exponent (known as a block-floating-point representation), which in effect turns the floating-point values into integers. 
A linear decorrelating transform similar to the \emph{discrete cosine transform} is then applied along each dimension (using only integer additions, subtractions, and bit shifts), followed by a JPEG-like zig-zag ordering of coefficients. 
Coefficients are then converted from two's complement to \emph{negabinary}---a base $-2$ representation of signed values---followed by a bitplane coding step that exploits the sparsity of decorrelated coefficients expressed in negabinary.
This encoding from most to least significant bit can be terminated at any point, for example, to satisfy a fixed storage budget or as soon as an error tolerance is met.

\subsubsection{Error Controls}

\zfp offers five compression modes: expert, fixed-rate, fixed-precision, fixed-accuracy, and reversible mode.
Expert mode allows fine-grained control of both accuracy and precision.
Fixed-rate mode uses a fixed number of bits per block.
Fixed-precision mode uses a fixed maximum number of bits per block.
Fixed-accuracy mode applies the error bound of $|x-\tilde{x}| \leq 2^\epsilon$ and can be used to implement an absolute error bound.
The reversible mode is lossless.

\subsubsection{Hardware Support}

Because of its data decomposition into small, independent blocks, \zfp supports massively parallel (de)compression via OpenMP, CUDA, and HIP backends.
And because of its simplicity and lack of data dependence (e.g., no dictionaries or probability models need to be learned), \zfp is among the fastest compressors available, achieving up to 1~TB/s throughput.
This makes \zfp suitable for batch compressing large datasets in parallel.
Importantly, when used as in-memory representation, \zfp array accesses can be made very fast, to the point where some applications see a net performance gain due to reduced data movement.
Compressed \zfp blocks are usually on the order of 1--2 hardware cache lines, while the corresponding decompressed data may occupy one to two orders of magnitude more space.
This makes hardware caching of compressed blocks an attractive solution to reducing data movement, where small ``nuggets'' of data are decompressed, processed, and then compressed to cache again, with only compulsory main memory accesses needed.

\subsubsection{Unique Features}

\zfp offers an alternative number format for multidimensional arrays that exhibit ``smoothness' or autocorrelation, as is the case with most fields representing physical quantities.
Compared with IEEE floating point and many recent variants, such as BFloat16, TensorFloats, Posits, and Blaz, \zfp provides much higher accuracy per bit stored. 
Error bounds can be specified, not just for a single application of \zfp compression, but also when \zfp is used in iterative methods, where compression errors may propagate and cascade.
Recent work has also provided mechanisms to ensure that \zfp compression errors are largely spatially independent, unbiased, and normally distributed, allowing applications to treat such errors as ``white noise.''
Additionally, \zfp supports fully lossless compression of IEEE floating-point arrays, as well as 32- and 64-bit integer data.

\zfp provides a common C++ array API to read and write individual array elements and is meant to serve as a substitute for conventional uncompressed array classes, such as STL vectors.
As such, \zfp arrays may be used wherever conventional arrays are used,  with minimal application code changes.
Unlike conventional arrays, however, \zfp provides the user with parameters both for setting an exact memory footprint or for error tolerance.
\zfp's array classes handle on-demand compression, decompression, and caching of recently accessed decompressed blocks of data, thus avoiding expensive (de)compress operations for each and every array access.
The fundamental unit of data in \zfp is a $d$-dimensional block of $4^d$ scalars (e.g., $4 \times 4 \times 4$ scalars in three dimensions), and such blocks are (de)compressed entirely independently of other blocks, possibly in parallel.
Although designed to limit memory footprint in numerical computations, \zfp also finds utility in reducing data movement, for example, between RAM and registers, between CPU and GPU, in communication between compute nodes, and when reading from or writing to disk.


\subsection{MGARD}
MGARD (\url{https://github.com/CODARcode/MGARD}) is a lossy compression framework built on finite-element analysis and wavelet theories. 

\subsubsection{Principles}
The key steps in MGARD are multilevel decomposition, quantization, and integer lossless encoding.

\textbf{Multilevel Decomposition}: The key to MGARD is the hierarchical decomposition algorithm~\cite{ainsworth2018multilevel, ainsworth2019multilevel}. Basically, MGARD treats data as a piecewise linear function on the initial grid and decomposes it in an iterative fashion using a predefined grid hierarchy. In each iteration, piecewise linear interpolation is used to approximate missing nodes (representing nodes absent in the next level) and then subtracted from their current values to obtain multilevel components. The multilevel components then  are mapped to the nodal nodes (representing nodes that exist at the current and next levels) to compute corrections using $L^2$ projection. Finally, the corrections are added to the current values of the nodal nodes to obtain the data representation in the next level. This procedure repeats until the coarsest grid is reached.

\textbf{Quantization}. Generally, MGARD uses linear-scaling quantization on multilevel components to enable error control on raw data and certain families of downstream quantities of interest. It also provides a nonuniform quantization scheme to better preserve features~\cite{gong2022region, gong2023spatiotemporally}.

\textbf{Integer Lossless Encoding}. MGARD applies integer lossless encoding in a similar way to SZ. Please refer to Section~\ref{compressor:sz:principles} for details.

\subsubsection{Error Controls}
MGARD features guaranteed error control on raw data and has unique features for downstream quantities of interest. 
MGARD supports error controls on common metrics such as $L^\infty$ errors and $L^2$ errors.
These bound the expressions $L_\infty = \max_i^N{\left|x_i-\tilde{x_i}\right|}$ and $L_2 = \sqrt{\sum_i^n\left(x_i-\tilde{x_i}\right)^2}$, respectively, with $L_\infty$ being equivalent to a pointwise absolute error bound.
It also provides error control on certain families of downstream quantities of interest with rigorous theories~\cite{ainsworth2019qoi}. 
Recently, this feature has been further enhanced by coupling with machine learning techniques~\cite{banerjee2022scalable, banerjee2023online, lee2022error}.
The decomposition and error control theories of MGARD extend to unstructured grids~\cite{ainsworth2020multilevel}, an area considered challenging for traditional compressors designed for structured grids.

\subsubsection{Hardware Support}

MGARD has been carefully optimized and engineered on both CPUs and GPUs~\cite{liang2021mgard+, chen2021accelerating, gong2023mgard}. 
It leverages platform portability and modern software engineering practice through tailored implementations with OpenMP, CUDA, HIP, and SYCL.
Specifically, it features unified APIs and memory buffers across CPUs and GPUs, self-describing data formats, and efficient out-of-core processing.

\subsubsection{Unique Features}

A primary way that MGARD distinguishes itself is its robust notions of error bounds for quantities of interest compared with other compressors and support for structured non-cartesian grids not offered by most other compressors.

Another novel featured MGARD offers is data refactoring and progressive retrieval~\cite{liang2021error}. 
This mode archives data nearly losslessly using multilevel decomposition and bitplane encoding and allows for on-demand data retrieval with error control in an incremental fashion. 
This has been further incorporated with erasure encoding to reduce storage and network overhead while maintaining data availability~\cite{wan2023rapids}.

\subsubsection{History and Impact}

MGARD, more than other compressors, emphasizes its strong mathematical heritage with an extensive line of papers proving new guarantees about the errors that can be preserved in derived quantities. This work began with univariate, then proceeded to multivariate preservation of errors.  It was extended to preserving derived quantities of interest starting with bounded linear functionals but eventually extending to some type of nonlinear functions giving robust proofs of its correctness on specific applications.

\subsection{LC}

LC (\url{https://github.com/burtscher/LC-framework/}) is a framework for automatically generating customized lossless and guaranteed-error-bounded lossy data compression algorithms for individual files or groups of files.

\subsubsection{Principles}
LC consists of three parts: a  romponent library, a  preprocessor library, and a  framework that combines them.

\begin{figure}[!htbp]
  \begin{center}
  \includegraphics[width=1\columnwidth]{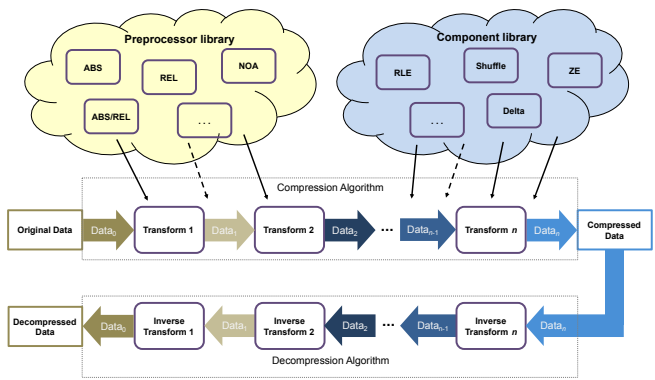}
  \end{center}
  \vspace*{-1em}
  \caption{LC's process of chaining (i.e., pipelining) $n$ data transformations to form a custom compression algorithm and the inverses of those transformations to form the matching decompression algorithm (the components are lossless whereas the preprocessors include guaranteed-error-bounded lossy quantizers).}
  \label{fig:LC}
\end{figure}

Both libraries contain data transformations (encoders) and their inverses (decoders) for CPU and GPU execution. The user can extend these libraries as explained in the tutorial. The framework takes preprocessors and components from these libraries and chains them into a pipeline to build a compression algorithm. It similarly chains the corresponding decoders in the opposite order to build the matching decompression algorithm. Figure~\ref{fig:LC} illustrates this process. Importantly, LC can automatically search for effective compression algorithms by testing all combinations of user-selected sets of components in each pipeline stage.

LC includes an extensive library of components and preprocessors. Most of them support 1-, 2-, 4-, and 8-byte word sizes. Both libraries are user customizable and extensible, meaning users are able to add their own data transformations by following the API outlined in the tutorial. LC then includes the new transformations in its search for a good compression algorithm and can use them in the code generator.

This focus on very low-level byte-level operations distinguishes it from frameworks such as SZ, which features more advanced notions of prediction to better decorrelate data in lossy compressors.

\subsubsection{Error Controls}
In addition to lossless algorithms, LC can generate lossy algorithms for 32-bit single and 64-bit double-precision floating-point data. It supports absolute, relative, normalized absolute, and combined absolute and relative error bounds. Moreover, it guarantees that these pointwise error bounds are not violated by losslessly encoding any value that it cannot quantize within the provided error bound. It supports all floating-point values, including infinities, NaN's, and denormals. Each quantizer provides two modes, one that replaces the lost bits by zeros and another that replaces them by random bits to minimize autocorrelation between the errors.

\subsubsection{Hardware Support}
LC can run on and generate algorithms for CPUs and GPUs. The algorithms are deterministic and fully compatible, meaning the user may compress a file on either the CPU or GPU and decompress the resulting file on either the CPU or GPU. The CPU code is written in C++ and parallelized using OpenMP. The GPU code is written in CUDA. Once a suitable algorithm has been found, the user can employ LC's code generator to produce a standalone compressor and decompressor for that algorithm that does not require the framework.
\subsubsection{Unique Features}
LC supports both exhaustive search for the best algorithm in the search space and a genetic-algorithm-based search for cases where the exhaustive search would take too long. In addition, the user can optionally supply a regular expression to reduce the size of the search space. LC is able to search for the best algorithm based solely on compression ratio or based on both compression ratio and throughput. In the latter case, it outputs the Pareto front, that is, a set of algorithms that represent different compression ratio versus speed trade-offs.

\subsubsection{History and Impact}

LC is a comparatively new compression framework.
It was initially designed as a component library for lossless compressors and later extended to include support for lossy compressors.
This history and focus offer very high levels of performance for lossless compression.

\subsection{SPERR}
SPERR~\cite{10177487} (\texttt{github.com/NCAR/SPERR}) 
is a wavelet-based compressor tailored for 
2D and 3D scientific data compression.

\subsubsection{Principles}
SPERR comprises three major data processing steps. 

\textbf{Wavelet Transform}: this step transforms the input data
  into wavelet \textit{coefficients} in the wavelet space, where 
  data is decorrelated and its information content is compacted 
  to a small number of large-magnitude coefficients.
  The vast majority of coefficients are very close to zero. 

\textbf{Coefficient Coding}: this step quantizes the floating-point
  wavelet coefficients into integers and encodes the integers bitplane
  by bitplane from the most significant ones to the least significant
  ones.
  The encoding algorithm, SPECK~\cite{1347192}, takes advantage of 
  the fact that the large-magnitude coefficients are sparse and are
  often clustered, achieving very high coding efficiency.

\textbf{\textit{(Optional) Outlier Correction}}: this step is designed
  for applications where a strict absolute error bound is required. 
  SPERR identifies all the \textit{outliers} whose error is beyond
  the prescribed error tolerance and encodes \textit{correctors} 
  that will bring the outliers back to the error 
  tolerance during decompression.
  The outliers most often account for a small percentage of the
  total number of data points.

\subsubsection{Error Controls}
SPERR supports three quality controls: (1) fixed size, (2) fixed 
peak signal-to-noise ratio, and (3) fixed maximum pointwise 
error, which is also referred to as fixed absolute error.
Internally, SPERR adjusts the quantization step size and
encoding termination conditions in the coefficient coding
step (step~2) to achieve the  prescribed compression quality.

\subsubsection{Hardware Support}
SPERR is implemented as  multicore CPU-based compressor and does not currently feature a GPU-based mode.

\subsubsection{Unique Features}
Compared with other established compressors, SPERR excels in compression
efficiency: SPERR most likely uses the least amount of storage to 
achieve a specific compression quality, often by a comfortable 
margin~\cite{10177487}.
At the same time, SPERR falls short in runtime performance, often by a 
factor of $\sim5X$ compared with the fastest performers.

What makes SPERR really stand out is its two special decoding modes:
 \textit{flexible-rate} and  \textit{multiresolution} decoding.

\textbf{Progressive flexible-rate decoding} means that any substring 
  of a compressed SPERR  bitstream, 
  given that it starts from the very beginning, is still valid
  for decompression, although the  reconstruction is of lower quality.
  This property is made possible by the  \textit{embedded} nature
  of compressed SPERR bitstreams.
  Flexible-rate decoding enables saving high-quality, large-volume data
  in a centralized repository and producing lower-quality, smaller-volume
  data with little cost (i.e., by truncating the compressed bitstream) 
  for downstream  applications with various quality-size trade-offs. 
  It also enables advanced data management such as \textit{tiered storage}, where 
  the smallest in volume but most frequently used portions of the 
  compressed bitstream are kept on hot storage and the bulk of the remainder bitstream for 
  the highest-quality reconstruction is kept on cold storage.

\textbf{Multiresolution decoding} means that in addition to the
  native resolution reconstruction, a hierarchy of the data with coarsened 
  resolutions is produced during decompression.
  This multiresolution hierarchy is enabled by wavelet transforms,
  which naturally approximate the input in multiple levels of
  lower resolutions.
  Compared with na\"ive multiresolution approaches such as
  sampling and subsetting, wavelets produce approximations of significantly higher 
  qualities and do not incur redundant storage.
  Multiresolution decoding enables data analysis under constraints
  (e.g., hardware capabilities and/or time) before devoting a significant
  amount of resources to a particular analysis routine.
  This approach is especially useful in
  exploratory workflows  such as scientific visualization.
  
We note that flexible-rate and multiresolution decoding
are both achieved with special controls during decompression;
in practice, \textit{all} compressed SPERR bitstreams support these 
two special decoding modes.

\subsubsection{History and Impact}

SPERR is a relatively newer compressor designed for climate data compression.
It builds on the existing SPECK encoding to achieve its multiresolution and flexible rate decoding features.

\subsection{DCTZ}
DCTZ (\url{https://github.com/swson/DCTZ}) is a transform-based lossy compressor inspired by discrete cosine transform (DCT), specifically DCT-II, and is designed to work with floating point  (single- or double-precision) in scientific and Internet of Things datasets.

\subsubsection{Principles}

\begin{figure}[h] 
\centering 
    \includegraphics[angle=90,width=\linewidth]{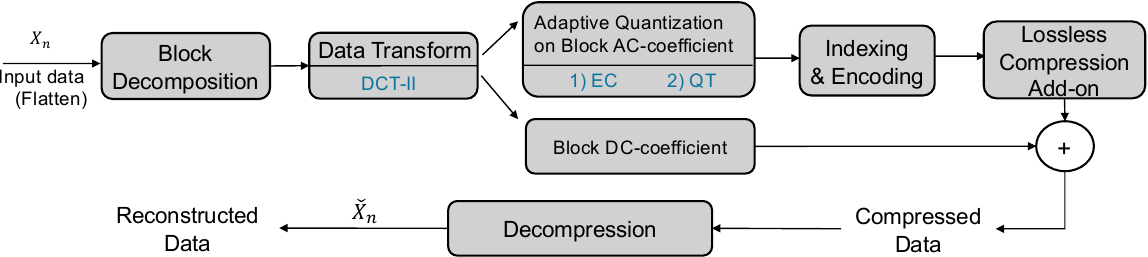} 
    \caption{Overview of DCTZ framework.}\label{fig:dctz-overview}
\end{figure}

Figure~\ref{fig:dctz-overview} shows how the current DCTZ compression framework works. DCTZ first decomposes the input data (flattened floating-point number array) into blocks with 64 data points each. Prior studies show that 8 by 8 metric can apply to floating-point numbers while providing a high energy compaction property~\cite{dctz-msst19, dctz-AC-normalization-HPEC20, dctz-precond-iwbdr22}. This block decomposition also helps improve overall compression and decompression performance. The next step is to apply DCT on each block to retrieve the DCT coefficients representing the input data in the frequency domain. Each block's first coefficient (DC) is saved as full precision to preserve the most crucial information. For the rest of the majority part coefficients (AC), DCTZ traverses every value to check whether they are inside the bin range. The AC coefficients inside the bin range will be quantized by using a uniform quantizer. This step will introduce compression error due to the truncation of the coefficients. The last step is to compress the data from previous steps with a lossless compressor such as zlib. DCTZ's decompression process follows the exact inverse step of the compression process. 

\subsubsection{Error Controls}
The critical step that affects compression performance in DCTZ is the quantization process, which maps a range of values, DCT coefficients in this case, to a small fixed one. Since most original data information is preserved in a few low-frequency coefficients, we store them as is and adopt a proper quantization technique on the high-frequency coefficients. The size of the bin range is decided by the number of bins ($B$) and the user-defined error bound ($P$), and then the bin range is defined as [-$P$*$B$, $P$*$B$]. This range is divided into ($B$) small ranges, each with a size of 2*$P$. As a result, the coefficients that fall inside the bin range will be mapped to an integer from 0 to 254 with the 1-byte bin index representation. An index of 255 is dedicated to DC or AC coefficients that must be saved as the original precision to preserve accuracy. During decompression, DCTZ uses the center value of each small bin range to represent the original value of the DCT coefficient.

We note that the binning mechanism described above is applied in the frequency domain (i.e., DCT coefficients), not in the spatial domain (i.e., original data). Therefore, extra errors could be introduced during the inverse transform to reconstruct data from a lossy state. If the maximum compression errors (the difference between reconstructed and original data) must be guaranteed within the user-specified error bound $P$, a revised error bounding method is needed. This strict error guarantee depends on the transform employed because each transform has a different inverse transform property. For DCT, its inverse transform has the same computation as the non-inverse one, calculated as the sum of weighted coefficients. Mathematically speaking, the new max error in the spatial domain is then calculated as $\sqrt{N}$ times the max error in the frequency domain (where $N$ is the block size). Therefore, users need to set their error bound to $P$/$\sqrt{N}$ in the frequency domain such that, after inverse transforming, the compression errors are bounded within $P$ in the original domain. This makes DCT with Quantizer-EC (DCT-EC) a conservative yet efficient compressor. In other words, it guarantees the user-defined error with a straightforward quantization process. 

\subsubsection{Hardware Suppport}

DCTZ is implemented as a serial CPU based compressor.

\subsubsection{Unique Features}
What distinguishes DCTZ from other compressors that use near-orthogonal transforms to decorrelate data is its quantization design.

Quantizer-EC applied the quantization to AC coefficients (high-frequency) directly. However, one can improve compression ratios further by applying various quantization methods to AC coefficients to reduce the number of bits required for encoding. This is inspired by the property of discrete transforms wherein spatial frequencies represent the detailed information of the original data. In other words, if the original data values are spatially smooth (common in many scientific applications that model physical phenomena or time-series IoT datasets), a block in the DCT domain will have smooth high-frequency coefficients (i.e., clustered with small variations).

Since most block coefficients show descent smoothness and repetitiveness, we design a quantization table $QT$ in our quantizer, Quantizer-QT. We generate $qt$ by finding the maximum value of the $n^{th}$ coefficient over all the partitioned blocks and build a quantization table of length $N-1$, where $N$ is the block size and $n\leq N$. Note that the DC coefficients of the blocks are not included in this step, since they are saved as is. $QT$ is calculated as $QT_{n,1}=\max\left \{ \left |  BA_{n,1}\right |, \left |BA_{n,2}\right |, \left |BA_{n,3}\right |,...,\left |BA_{n,m} \right |\right \}$, where $m$ is the total number of decomposed blocks and the input data is a one-dimensional floating-point array. Then, all AC coefficients are converted into a global bound and quantized by using Quantizer-EC after being divided by $QT$. 

\subsubsection{History and Impact}

DCTZ is a newer compressor introduced in 2019~\cite{dctz-msst19, dctz-AC-normalization-HPEC20, dctz-precond-iwbdr22, dctz-characterization-drbsd22}.
It has demonstrated that it can achieve high compression ratios while guaranteeing specific error bounds and comparable performance with SZ and ZFP~\cite{dctz-precond-iwbdr22}.

\subsection{TEZip}

TEZip  (\url{https://tezip.readthedocs.io/}) or Time Evolutionary Zip is developed in RIKEN R-CCS and designed to compress time
evolutionary data by using deep learning for prediction.

\subsubsection{Principles}
The TEZip compression/decompression
procedure consists of three steps: model training, compression, and decompression using a deep learning approach. Specifically, TEZip uses PredNet to predict future frames to maximize the compression ratio of image and video data.

\begin{figure}[h] 
    \centering 
    \includegraphics[width=\linewidth]{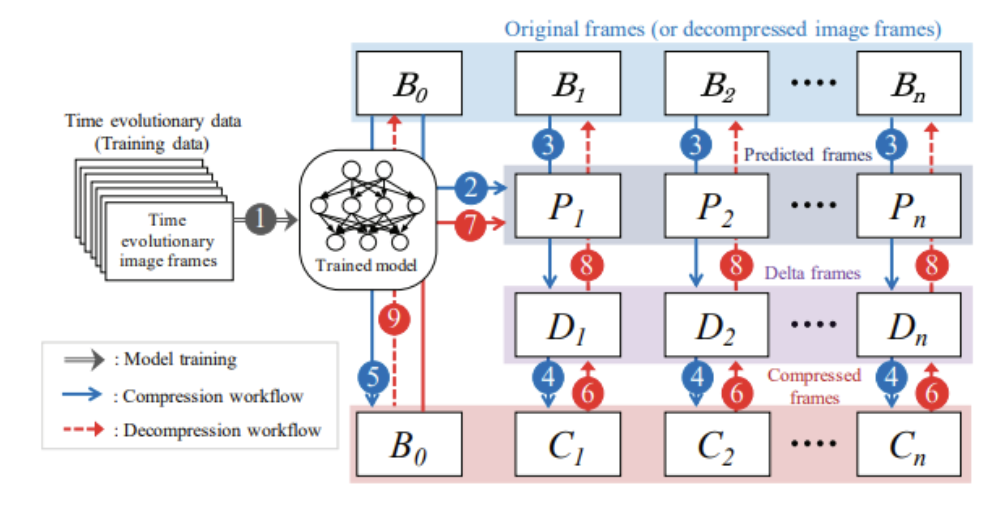} 
    \caption{{TEZip (Time Evolutionary Zip) framework}
}
    \label{fig:tezip} 
\end{figure}

Figure~\ref{fig:tezip} shows how TEZip leverages the time evolutionary image frames as training data. The trained model predicts future frames (denoted as $P_1, P_2, \ldots, P_n$) from original frames ($B_0, B_1, \ldots, B_n$). The compression workflow (blue arrows) calculates delta frames ($D_1, D_2, \ldots, D_n$) as the difference between the original and predicted frames, which are then compressed into $C_1, C_2, \ldots, C_n$. The decompression workflow (red dashed arrows) reconstructs the original frames from the compressed data by reversing the compression process, utilizing the delta frames and the trained model's predictions.

The TEZip \cite{Roy2021} approach ensures that the system effectively learns, compresses, and restores time evolutionary data, achieving compression rates while maintaining the quality of the restored images.

\textbf{Learning}: The TEZip framework utilizes a prediction model to learn the temporal sequences of objects over time. The input  data is converted into the .hkl format and used for training the model. This training process enables the model to predict future frames based on past observations, capturing the dynamics of the object movements.

\textbf{Compression}: The trained prediction model is then utilized in the compression process. This involves compressing the inference results and the differences between the time evolutionary images. By calculating the difference between the original image and the inference result, various encoding methods are applied. These processes effectively increase the compression rate by minimizing the amount of data required to represent the temporal changes.

\textbf{Decompression}: Using the trained model and the binary file (.dat) generated during the compression process, the original sequence of images can be restored. Keyframes are input to reproduce the results of the compression process. The decoding process, including density-based spatial decoding and partitioned entropy decoding, is executed in reverse order to recover the original differences. The error-bounded quantization process, being a lossy compression technique, is excluded from the decompression. The original images then are restored and output by combining the inference results with the recovered differences.

\subsubsection{Error Controls}

TEZip uses PredNet (prediction) architecture, a convolutional LSTM model that predicts the instrumental image frames based on past data. To enhance the accuracy of inferences, TEZip implements initial frames of a new sequence using the final states of the preceding sequence. The ``warm-up" process utilizes the temporal continuity between consecutive image segments that are stabilizing the network state and improving the prediction accuracy for subsequent frames.

\subsubsection{Hardware Support}
Older versions of TEZip are implemented by using TensorFlow whereas newer versions are being implemented in PyTorch.  The compressor can be used on hardware platforms that these libraries support.

\subsubsection{Unique Features}

Inferring subsequent frames progressively from previous inference results could lead to  gradual degradation in image accuracy.
TEZip is uniquely optimized for time-series-based compression and utilizes the notion of key frames often used in video compression formats.
In order to mitigate and maintain a level of accuracy, it is essential to periodically incorporate inference based on the original image data. This can be achieved through two approaches: Static Window-based Prediction (SWP) and Dynamic Window-based Prediction(DWP). TEZip use both approaches that stabilize the inference quality by adjusting the prediction window that more reliable continuity in the image sequence. 

\textbf{Static Window-based Prediction (SWP)}: A fixed value determines the number of frames to be predicted from a single image. After specified number of frames has been generated, the model uses the last predicted frame as the keyframe to infer the next set of frames. This is repeated for all images.

\textbf{Dynamic Window-based Prediction (DWP)}: The MSE (mean squared error) is calculated between the original image and predicted. If the MSE remain the below the present threshold, the inference process proceeds with the current keyframe. If the MSE exceeds the threshold, however, the keyframe is updated to the most recent image from which the prediction meets the required accuracy. This cycle is repeated for each image in the sequence, ensuring that the prediction quality is maintained throughout the process. 

\subsubsection{History and Impact}

TEZip is a comparatively newer compressor.  It can achieve very high compression ratios albeit at very low throughput and requires a learning process on a similar dataset for peak effectiveness.


\subsection{LibPressio}

LibPressio(\url{https://github.com/robertu94/libpressio}) is not a compressor itself, but it provides a common, lightweight interface to many compressors including all the compressors listed above.

\subsubsection{Principles}
LibPressio aims to be a low-overhead abstraction that provides common interfaces for common tasks while not restricting more advanced use cases where they are supported.  As such, it relies on the underlying compressors to perform compression.

\subsubsection{Error Controls}
LibPressio does not provide error controls of its own, but it provides several other features: (1) standardized names for common error bound modes such as pointwise absolute error bounds, and psnr; (2) mechanisms to use any uncommon or highly specialized error-bounded mode supported by the underling compressors, for example, MGARD's QOI modes that preserve bounded linear functionals not widely supported by other compressors; and (3) a module that allows translating one type of common error-bounded mode supported by various compressors to others (e.g., absolute to value-range relative or absolute to pointwise relative).

\subsubsection{Hardware Support}
LibPressio provides sophisticated support for compression on CPUs and GPUs by automatically facilitating advanced optimizations such as GPU direct on platforms that support it and migration of data between CPU- and GPU-based phases of compression, allowing users to quickly experiment with different compression pipelines with minimal changes to their application codes.
However, the hardware support features are user-extensible for other platforms using its domains feature, which enables data migrations between heterogeneous memories and enables third-party support for additional hardware devices.

\subsubsection{Unique Features}

LibPressio provides several features critical to the adoption of compressors in scientific codes.
(1)\textit{Bridge between Compressors and Applications} The aim here is that each can evolve independently even as compressors change dramatically.  For applications, LibPressio provides a consistent API and standardizes the naming of error bounds and introspection capabilities to enable programmatic discovery and use of new compressors.  For compressors, LibPressio standardizes a way for applications to provide data quality metrics to compressors to enable automated tuning, validation, and similar use cases.
(2) \textit{Efficient Exposure of Capabilities of All Supported Compressors } In addition to standard options, compressors can provide their specialized options in a name-spaced way that can be introspected by applications to allow users all of the underlying capabilities of the compression library so that they do not need to reach for lower-level functions from the compression libraries.
(3) \textit{Debugging of Compression Pipelines} In addition to providing data quality metrics, compressors can export internal counters, metrics, and views of intermediate data to enable robust visualization and analysis as compression runs in real time.  These features can be implemented without code changes in applications, enabling debugging on demand with minimal effort.
(4) \textit{Common and Efficient Implementations for I/O Library Extensions and Programming Language Bindings} This feature means that each compressor does not need to develop these capabilities separately.  Applications can easily adopt compression techniques regardless of their I/O library or programming language. Compressors benefit from dramatically reduced development and maintenance costs.
(5) \textit{Generic Implementations of Compressor Features} Generic implmentations mean that these features do not need to be implemented separately for every compression library.  For example (i) embedding of provenance and configuration metadata in the compressed stream to improve portability, (ii) automatic configuration of compressors to meet user-specified error bounds to simplify configuration \cite{underwood_optzconfig_2022}, (iii) automatic CPU parallelization of thread-safe compression libraries to improve utilization and performance, (iv) prediction of compression performance with minimal recomputation of metrics, (v) preprocessing-based techniques to convert an absolute error bound to a pointwise relative error bound, and (vi)) standard configuration file formats.
(6) \textit{Loading of New Compressors without Recompilation} LibPressio supports dynamically loading new compressors by linking them into an application allowing easy experimentation and development of license-incompatible\footnote{LibPressio uses a BSD license, so compression modules that are under GPL licences do not ``infect'' the rest of the code base.} or closed-source modules\footnote{For example, during development compressors are often closed source until the paper is published.  In other cases, sponsors of compressors may require a period of exclusivity.} without forking or modifying LibPressio.

\subsubsection{History and Impact}

LibPressio has been used widely in 6 U.S. DOE labs, 2 international super-computing centers, and 7 universities.  It has over 250 unique monthly downloads from GitHub.  It has been integrated into Spack and Anaconda, enabling ease of adoption.

\section{Gap Analysis}
Across many domains and applications, error-bounded lossy compression techniques are increasingly important aspects of workflows to provide additional storage capacity, improve throughput, or even increase memory capacity.  We present here a high-level analysis of the findings of the needs of applications and where compression technologies can improve to meet those needs.

\subsection{Use Cases for Compression} The majority of applications considering compression are doing so to save storage (7 of 9) or throughput (5 of 9), with slightly fewer applications looking to improve throughput. Of the 9 applications,  7 report that a compression ratio of at least 5 is required to adopt compression, and many described this requirement as an improvement over lossless compression rather than no compression.  Likewise, applications that describe throughput as a priority want to see application speedups that exceed what they can achieve with lossless compression; 2 of the 9 applications want to see compression helping them meet a real-time streaming bandwidth target.
This data highlights that applications need more than just modest improvements to adopt compression because in many cases it represents an increase in the complexity of their workflows and software deployments.  

Of the studied workflows, 6 of the 9 applications want to perform compression on the CPU, a number that is expected to decrease to 4 out of 9,
while 5 of the 9 applications want to compress on the GPU, a number that is expected to increase to 7 out of 9 in the near future.  This shift largely represents the shift of applications from CPUs to GPUs to leverage the GPU capacity on leadership-class computing facilities.
One application group---light sources---called out the use of FPGAs as also increasingly important for their application use cases.
FPGAs already
are used in these applications, so incorporating their use for compression is consistent with other work in the field.
One critical pair application that was not included in this report---AI training and inference---uses other forms of specialized hardware in addition to GPUs, such as TPUs and Cerebras wafer-scale engines.  We intend to study use cases of compression in these applications in a future version of this report.

Of the 9 applications, 3
describe the need for interoperable compression and decompression on different hardware platforms; so far only two compressors fully meet this requirement, and only one implements byte-for-byte interoperability.  In some cases, this kind of interoperability can be difficult to efficiently implement (e.g., Huffman tree construction on a GPU \cite{tian_revisiting_2021}), difficult to implement correctly because of platform differences (e.g., LC reimplemented core math function on the GPU to ensure byte-for-byte interoperability with the CPU),  or difficult to implement at all on all platforms because of platform limitations (e.g., lack of global memory in Cerebras requiring an alternative Huffman tree implementation \cite{ceresz}).   In other cases, there is substantial difficulty in supporting multiple platforms with the same codebase, but that is improving (e.g., with the recent version of cuSZ and MGARD) with the adoption of performance portability libraries and designs.
\subsection{Quantities of Interest}.  One area where compressors can improve is the preservation of higher-order quantities of interest.  Of the 9 applications surveyed, all but 1 indicated that they found it difficult to preserve their quantities of interest with existing production compressors.  Three major groups of quantities of interest need additional focus by compression developers: derived quantities of interest, topological features, and distributional features.

    \textit{Derived quantities of interest} are scalar values derived with an explicit formula from the data or its error (e.g., dSSIM, descriptive statistics).  At least 4 of the 9 applications have at least one of these that need to be preserved.  While in many cases a relationship exists between the error bound and the derived QoI (see \cite{tao_fixed-psnr_2018} for an early example), it is nontrivial to explicitly derive this relationship.  Some work in this area has been done \cite{banerjee2022scalable, liu_qpet_2024}, but these techniques are either difficult to use, still requiring extensive mathematical proofs to establish correctness, or are not fully integrated into production compressors adopted by applications. Moreover, if one can derive the relationships between application-derived QoI, the bound may be very pessimistic, resulting in lower than otherwise required compression ratios \cite{underwood_optzconfig_2022} c), and the run performance of the approach may be unacceptably low \cite{underwood_optzconfig_2022} for applications with large datasets.  More work is needed on both theory and application to make these techniques approachable to the applications that need them.
    
    \textit{Topological features} refer to the minima, maxima, and critical points that exist for data and its integrals or derivatives. At least 3 of the 9 applications cite a need to preserve these kinds of QoIs. While compressors exist for these types of bounds as well, they are largely research prototypes with high overheads and lacking integrations into appropriate libraries and languages where applications would use them \cite{msz} or they overpreserve the data by preserving all derivatives as part of preserving the Sobolev norm of data \cite{mgard}, resulting in lower than required performance.  The accessibility of functionality aspect can be improved by integration of existing or development of new research prototypes of compressors using frameworks such as LibPressio that export these functions automatically, but resource utilization improvements come from both algorithmic improvements and making production-ready the relevant codes.
    
    \textit{Distributional features} refer to the shape of the distribution of values either in some window or globally.  At least 2 of the 9 of applications cite a need to preserve these kinds of features either in the data itself or in decompression errors.  While the distribution of error bounds of compressors has been studied and characterized \cite{lindstrom_error_2017}, this work is substantially out of date compared with current compressors and is merely descriptive. Applications need proscriptive protection of the distribution of data values and errors, which is not supported by any major and possibly any research-grade compressor.

Another key aspect of the adoption of compressors it the simplicity by which applications can specify their quantities of interest and identify configurations of compressors that can meet their requirements.  Configure search tools such as OptZConfig \cite{underwood_optzconfig_2022} included with LibPressio can help with this process, but the overhead of these methods can still be very high \cite{rahman2023feature}, and the tools are not scalable to applications with a very large number of fields that potentially need to be configured differently.  More work is needed to help address this level of overhead.

In short, extensive work is needed to improve the performance, accessibility, and applicability of techniques to preserve higher-level quantities of interest to meet the needs of applications.

\subsection{Longevity of Compressed Data}
While 44\% of applications cite a need for only ephemeral compression---that is, as part of the workflow of an application and discarded afterward---the remaining 55\% of applications need long-term stability and support of the format of their compressed data to facilitate adoption.  The most common duration cited was at least 5--10 years, if not longer.  However, all the existing compressors are supported only by shorter-term funding, presenting a key challenge for the adoption of these methods for many applications.

\subsection{Mechanisms and Installation}

Advanced compressors are most useful when they support the languages and platforms used by applications.

Nearly all the applications, 8 of 9, use Python somewhere in their data analysis stack, so integration with Python is critical to the adoption of compressors; and only 3 of the studied families of compressors have Python bindings and Python packaging.  With LibPressio, the availability of Python bindings extends to 100\% of compressors with LibPressio bindings, but it does not automatically improve the packaging of compressors.  The LibPressio maintainers make an effort to ensure that all supported compressors are installable via spack\cite{gamblin_spack_2015}, but this represents only 44\% of applications.  Support in the Python packaging ecosystem for native libraries, especially around large complex C++ dependencies and GPU libraries, is lacking \cite{ralfgommersPypackagingnative2022}, making it difficult to support a large, complex native ecosystem.  A large number of applications, 55\%, still compile all their dependencies from source as part of a manual installation process, further limiting the adoption of any dependencies including compressors.

Moreover,
5 of the 9 application areas also utilize a lower-level language as a key component of their software stack.  Of these 5 applications, 4 use C++ and 1 uses Fortran90.  Fortran 2003 added minimal support for variable-length strings and c-style pointers, making it possible but significantly complicated for compressors to support Fortran.  Having individual compressors add support for Fortran 2003 or later via LibPressio is possible but would require substantial effort. For Fortran90, however, lacking this minimal support means it has no practical path to direct integration of compressors in a general way without resorting to nonstandard compiler extensions or the adoption of I/O libraries for Fortran that support compression, such as HDF5; and even this approach may not be possible given the subset of features of HDF5 available in Fortran90 \cite{thehdfgroupNewFeaturesHDF52011}.

In short, more work is needed to ensure that new compressor features are incorporated into tools that applications can use to meeting their compression objectives.

\subsection{Specialized Compression Needs}

Most of the applications,
8 of 9, were able to identify special needs that are not well served by current production compressors.
In this section we group and describe these needs.

The need for greater support for data structures was cited by
3 of the 9 applications.
Of these,
2 needed support for uncorrelated dimensions passed as a dense tensor. Without this feature, the compression ratio of prediction-based and transform-based compressors is unduly hampered by trying to relate unrelated elements of data stored in a dense tensor.  The research compressor CLIz \cite{cliz} supports this feature, but it has not been adopted by major compressors.

Support for unstructued gid data was cited as a need by
1 application.
Some research grade compressors do support this feature \cite{Ren_Liang_Guo_2024}, but it too is not widely adopted and is not supported by higher-level abstractions such as LibPressio.  Without this support, compressors have to treat this data as one-dimensional, which can dramatically limit the correlations that compressors can correctly leverage to preserve quality and increase compression ratios.

A requirement cited by 1 application was support for compression of heterogeneous columns of data streamed over a network (i.e., streaming dataframes).
While supported by some data-processing frameworks \cite{lentner_shared_2019}, these frameworks do not include support for modern lossy compression and need further study.

Of the 9 applications, 2 cited the need for 
support for additional operations on compressed data.
Another need cited by
1 application (system logs) was the ability to perform queries (\`a la SQL); no current compressor supports this operation efficiently.  A promising direction for this work is holomorphic compression \cite{hoszp}, which is an open and active research area in lossy compression; but since this is a newly identified need for applications, it requires further study.

Multiple applications reported needing support for random access decompression by block.  SZ2,\footnote{but not SZ3 because the required internal functions are not exported} ZFP, and SPERR include an API for these functions, but they are low-level and do not feature by higher-level abstractions in LibPressio that would work between compressors.  LibPressio has functionality that can be used to implement a similar function generically but with a size and in some cases runtime overhead compared with compressors that support this function natively.

A third of the applications, 
3 of 9, need greater optimizations to achieve bandwidth requirements during streaming.
Light Sources need careful co-design between the compressor and the data reduction pipeline infrastructure to meet bandwidth requirements with available hardware, including optimizations to streaming, GPU kernel launches, and compression algorithms to meet hardware requirements.
Fusion and system logs applications also report the need to support streaming of data to alleviate bandwidth requirements.
Streaming differs from traditional compression tasks in that the entire data is not available at once, meaning that decisions need to be made to balance throughput and compression ratios, an area  requiring further study.

In short, despite the nearly 20 years of research on modern error-bounded lossy compressors starting with fpzip \cite{lindstrom_fast_2006}, a steady stream of new use cases need to be identified, supported, and standardized for use by applications to support the needs of applications as they evolve.

\section{Conclusion}
This report presents the most comprehensive study of application needs for lossy compression developed to date.
We intend to continue to revise and prepare new reports to capture the ongoing development of applications for lossy compression and the development of compressors to serve those applications.
We note that while applications can largely find compressors on the platforms where needed, more work is required in order to ensure portability and to support the sophisticated error controls demanded by applications and their advanced use cases.
We also identify key barriers to adoption in the longevity of compression formats and the support for easy of installation/use in the Python ecosystem.
Further,
we identify a number of new specialized compression needs in applications as they grow and evolve.

\section*{Acknowledgments}
This research was supported by the Exascale Computing Project (ECP), Project Number: 17-SC-20-SC, a collaborative effort of two DOE organizations -- the Office of Science and the National Nuclear Security Administration, responsible for the planning and preparation of a capable exascale ecosystem, including software, applications, hardware, advanced system engineering, and early testbed platforms, to support the nation’s exascale computing imperative. The material was supported by the U.S. Department of Energy, Office of Science, Advanced Scientific Computing Research (ASCR), under contract DE-AC02-06CH11357, and supported by the National Science Foundation under Grant OAC-2003709/2303064, \\OAC-2104023/2247080, OAC-2311875/2311876/2311877,\\ OAC-2312673,  OAC-2034169, OAC-1751143, OAC-2330367, OAC-2313122, OAC-2311756, OIA-2327266 and OAC-2103621. We acknowledge the computing resources provided on Bebop (operated by the Laboratory Computing Resource Center at Argonne). Some of the experiments presented in this paper were carried out using the PlaFRIM experimental testbed, supported by Inria, CNRS (LABRI and IMB), Universit\'e de Bordeaux, Bordeaux INP and Conseil R\'egio\-nal d’Aquitaine (see https://www.plafrim.fr). 
TEZip work has been supported by the COE research grant in computational science from Hyogo Prefecture and Kobe City through the Foundation for Computational Science. XIOS-SZ - Mario Acosta and Xavier Yepes-Arbós have received co-funding from the State Research Agency through OEMES (PID2020-116324RA-I00). We thank the Texas Advanced Computing Center (TACC) at the University of Texas at Austin for providing computational resources on `Frontera' system \cite{Frontera2020}.
Use of the Linac Coherent Light Source (LCLS), SLAC National Accelerator Laboratory, is supported by the U.S. Department of Energy, Office of Science, Office of Basic Energy Sciences under Contract No. DEAC02-76SF00515.

This work has been supported in part by the Department of Energy, Office of Science under Award Number DE-SC0022223 as well as by equipment donations from NVIDIA Corporation.

This work has been co-funded by the European Union through ‘MDDB: Molecular Dynamics Data Bank. The European Repository for Biosimulation Data’ [101094651], and
the Swedish e-Science Research Center.

\bibliographystyle{abbrv}
\bibliography{ref}

\appendix
\section{Summary of Application Needs}
\begin{landscape}
\newcommand{\TwoLine}[3][l]{\begin{tabular}{@{}#1@{}}#2\\[-.6ex]#3\end{tabular}}

\begin{table}

	\caption{Summary of Application Requirements}\label{tab:requirements}

	\centering\renewcommand{\arraystretch}{1.3}

	\begin{tabularx}{\linewidth}{|X||XXlXXXXXXX|}
		\hline
		\textbf{Application}                   & \textbf{Needs}                          & \textbf{Device}                     & \textbf{Target CR}          & \textbf{Target Bandwidth}                       & \textbf{Analysis/QoI}                          & \textbf{Longevity}                & \textbf{Mechanism}                       & \textbf{Installation}                   & \textbf{Special needs}                   & \textbf{Format}                                \\
		\hline
		Molecular Dynamics & storage                        & CPU                        & $>3$               & $>1\times$                            & bonds and sequences              & 10 Years                 & C,\,C++, HDF5                      & Spack, Manual                  & Random Block Access             & Particles                             \\
		\hline
		Climate                       & storage                        & CPU~$\to$~GPU      & $2-3\times l$       & $>1\times l$                          & ldcpy                             & indefinite               & Python, Julia, R,\,HDF5, pnetcdf & site modules, pip/conda, spack & Uncorrelated Dims               & Dense\,$\to$ Unstructured Grid \\
		\hline
		Light Sources                 & throughput +storage & CPU\,$\to$ GPU, FPGA & $>10\times$        & real-time 1~TB/s                        & per beamline; manual\,$\to$ automatic   & 10 Years                 & Python, C++                     & conda\,$\to$ spack      & Uncorrelated Dims               & Dense                                 \\
		\hline
		Cosmology                     & storage +throughput & GPU                        & $>10\times$        & $>1\times l$                          & halo              & To be determined & HDF5, C++, python               & Manual                         & Hardware Portable Decompression & Dense                                 \\
		\hline
		Seismology                    & throughput +storage & GPU, CPU                    & $>20\times$        & $>1\times l$                          & visualization of the stacking image              & Ephmerial & Fortran90, CUDA, Python         & Manual                         & Asynchronous Batching           & Dense                                 \\
		\hline
		Combustion                    & storage                        & GPU                        & $2\times-5\times$  & not urgent                            & topological descriptors           & 5-10 years               & C++, Python, HDF5               & Manual, pip                    & High accuracy for feature preservation                                & Dense                                 \\
		\hline
		Fusion                        & storage +throughput & GPU                        & $>5\times$         & not\,urgent $\to$ $>1\times l$ & Spikes/peaks              & Ephmerial                & HDF5\,$\to$ Python       & Manual, site modules, pip      & streaming, provence             & 2D Dense                              \\
		\hline
		Quantum Circuit               & memory capacity     & GPU, CPU                    & $2\times-10\times$ & real-time 25~GB/s                       & conservation laws, distributional & Ephmerial                & Python                          & pip                            &                                 & 20-30D Dense                          \\
		\hline
		System Logs                   & throughput                     & CPU                        & $>10\times$        & $>10\times l$                         & 
        aggregate stats, anomalies, data for scheduling algorithms & Ephemeral                & Python                          & pip, conda, spack                      & Queries (\`a la SQL)            & 2D Tables                             \\
		\hline
	\end{tabularx}

\end{table}
\end{landscape}

\end{document}